\documentclass[a4paper,traditabstract]{aa} 
\usepackage{graphicx}
\usepackage{natbib}
\usepackage{txfonts}
\usepackage{longtable,lscape}

\newcommand{\Log}{\mbox{Log}}

\newcommand{\xmm}{XMM-{\em Newton}}
\newcommand{\chandra}{{\em Chandra}}

\newcommand{\lognlogs}{Log~$N$--Log~$S$}

\newcommand{\ergs}{erg s$^{-1}$}
\newcommand{\ergscmq}{erg s$^{-1}$ cm$^{-2}$}

\newcommand{\ergsHz}{erg s$^{-1}$ Hz$^{-1}$}

\newcommand{\e}[1]{\cdot 10^{#1}}

\newcommand{\sradio}{{S_{1.4 \mathrm GHz}}}

\begin{document}

   \title{X-ray properties of radio-selected star forming galaxies\\in the \chandra-COSMOS survey}

   \author{P. Ranalli
          \inst{1,2,3}
          \and
          A. Comastri\inst{3}
          \and
          G. Zamorani\inst{3}
          \and
          N. Cappelluti\inst{3}
          \and
          F. Civano\inst{4}
          \and
          I. Georgantopoulos\inst{2,3}
          \and
          R. Gilli\inst{3}
          \and
          E. Schinnerer \inst{5}
          \and
          V. Smol{\v c}i{\'c}\inst{6,7,8,9}
          \and
          C. Vignali\inst{1}
          }

   \institute{Universit\`a di Bologna, Dipartimento di Astronomia,
     via Ranzani 1, 40127 Bologna, Italy\\
     \email{piero.ranalli@oabo.inaf.it} 
    \and
     Institute of Astronomy and Astrophysics, National Observatory of
     Athens, Palaia Penteli, 15236 Athens, Greece
    \and
     INAF -- Osservatorio Astronomico di Bologna,
     via Ranzani 1, 40127 Bologna, Italy
    \and
     Harvard-Smithsonian Center for Astrophysics, 60 Garden Street,
     Cambridge, MA, 02138, USA
    \and
     Max-Planck Institut f\"ur Astronomie, K\"onigstuhl 17,
     69117 Heidelberg, Germany
    \and
     ESO ALMA COFUND Fellow
    \and
     Argelander Institut for Astronomy, Auf dem H\"ugel 71, 53121 Bonn,
     Germany
    \and
     European Southern Observatory, Karl-Schwartzschild-Stra{\ss}e 2,
     85748 Garching b.\ M\"unchen, Germany
    \and
     University of Zagreb, Physics Department, Bijeni\v{c}ka cesta 32, 10002 Zagreb, Croatia
}

   \date{Received 2011-12-22; accepted 2012-4-20}

   \abstract{
     X-ray surveys contain sizable numbers of star forming galaxies,
     beyond the AGN which usually make the majority of
     detections. Many methods to separate the two populations are used
     in the literature, based on X-ray and multiwavelength properties.
     We aim at a detailed test of the classification schemes and to study the
     X-ray properties of the resulting samples.

     We build on a sample of galaxies selected at 1.4 GHz in the
     VLA-COSMOS survey, classified  by
     \citeauthor{smolcic08}\ (2008) according to their optical colours
     and observed with \chandra.  A similarly selected control sample
     of AGN is also used for comparison. We review some X-ray based
     classification criteria and check how they affect the sample
     composition.
     The efficiency of the classification scheme devised by
     \citeauthor{smolcic08}\ (2008) is such that $\sim 30\%$ of
     composite/misclassified objects are expected because of the
     higher X-ray brightness of AGN with respect to galaxies.
     The latter fraction is actually $50\%$ in the X-ray detected
     sources, while it is expected to be much lower among X-ray undetected
     sources. Indeed, the analysis of the stacked spectrum of undetected
     sources shows, consistently, strongly different properties
     between the AGN and galaxy samples.  X-ray based selection
     criteria are then used to refine both samples.

     The radio/X-ray luminosity correlation for star forming galaxies
     is found to hold with the same X-ray/radio ratio valid for nearby
     galaxies. Some evolution of the ratio may be possible for sources
     at high redshift or high luminosity, tough it is likely explained
     by a bias arising from the radio selection.
     Finally, we discuss the X-ray number counts of star forming galaxies from
     the VLA- and C-COSMOS surveys according to different
       selection criteria, and compare them to the similar
     determination from the Chandra Deep Fields.
     The classification scheme proposed here may find application in
     future works and surveys.
}

   \keywords{ X-rays: galaxies -- radio continuum: galaxies --
     galaxies: fundamental parameters -- galaxies: active 
     -- galaxies: high redshift
   }

   \maketitle

\section{Introduction}

Radio and far-infrared observations have been widely accepted as
unbiased estimators of star formation (SF) in spiral galaxies for
decades \citep[see the][reviews]{cond92,kenn98}. The X-ray domain has
also been recognized as a SF tracer in non-active galaxies (hereafter
just ``galaxies'') thanks to a number of works highlighting the
presence of X-ray vs.\ radio/infrared correlations
(\citealt{djf92,grimm02}; \citealt{rcs03}, hereafter RCS03;
\citealt{gilfanov04a}).  Strong absorption (i.e.\ with column
densities $\gtrsim 10^{22}$ cm$^{-2}$) is also rare among galaxies,
making the X-ray domain scarcely sensitive to extinction. Thus, an
X-ray based Star Formation Rate (SFR) indicator can be considered not
biased by absorption (RCS03).  An
interpretation framework, whose main idea is the dominance of
High-Mass X-ray Binaries among the contributors to the X-ray
luminosity of galaxies, has also been developed
\citep{gilfanov04b,pr07} and is currently the subject of further
investigation. The observations of deep fields, especially with
\chandra, have prompted the search for galaxies at high redshifts
\citep{alexander02,bauer02,hornsch03,ranalli03AN}. The galaxies X-ray
luminosity function and its evolution has been investigated both in
the local universe and at high redshift
(\citealt{ioannis99,norman04,georgantopoulos05}; \citealt[][hereafter
RCS05]{rcs05}; \citealt{georgakakis07,ptak07,lehmer08})
with also the goals of obtaining an absorption-free estimate of the
cosmic star formation history, and deriving the contribution by
galaxies to the X-ray background.

However, any work involving galaxies in X-ray surveys has to deal with
the fundamental fact that AGN are preferentially selected in
flux-limited X-ray surveys. A careful and efficient classification of
the detected objects is necessary to identify the galaxies among the
dominant AGN population.   In early studies
  \citep{maccacaro88} AGN were found to populate a well defined region
  of the X-ray/optical vs.\ X-ray flux plane, bounded by an
  X-ray/optical flux ratio $X/O=-1$ (see definition in
  Sect.~\ref{sec:X/O}). This threshold has been often adopted as an
  approximate line dividing AGN and X-ray emitting galaxies. A more
  robust separation between AGN and star forming galaxies is obtained
  \citep{xue-cdfs4Ms,vattakunnel12} by considering several different
  criteria (X-ray hardness ratio, X-ray luminosity, optical
  spectroscopy, X-ray/infrared or X-ray/radio flux ratios). An
  analysis of the relative merits of the different criteria when taken
  separately, and of the most effective trade-offs to identify
  star-forming galaxies is one of the aims of this paper.

A similar need of a careful object classification has arisen in deep radio
observations. It has been known since long that a population of faint
radio sources associated with faint blue galaxies was emerging at
radio fluxes below $\sim 1$ mJy
\citep{windhorst85,fomalont91,vla98,vla00}. In recent years, it has
been shown that a sizable fraction (about 50\%) of this sub-mJy
population is actually made up of AGN
\citep{gruppioni99,ciliegi03,seymour08,smolcic08,vla-cdfs-padovani09,strazzullo10}. This
means that an accurate screening is needed also for radio-selected
faint galaxies.

This screening has been the subject of the work by \citet[][hereafter
S08]{smolcic08}, who made use of the extensive data sets of the COSMOS
survey to analyze $\sim 2400$ sub-mJy radio sources and classified
them according to a newly developed, photometry-based method to
separate SF galaxies and AGN. Their method is based on optical
rest-frame synthetic colours, which are the result of a principal
component analysis of many combinations of narrow-band colours, and
which correlate with the position of the objects in the classical BPT
diagram (\citealt{baldwin81}; see Sect.~\ref{sec:selezione} for
details).

Here we build on this work, and use the S08
samples as the starting point for our classification of the X-ray
galaxies in COSMOS. We intend to test the X-ray based selection criteria
against the S08 method, and eventually refine the
selection.

The Cosmological Evolution Survey (COSMOS) is an all-wavelength
survey, from radio to X-ray, designed to probe the formation and
evolution of astronomical objects as a function of cosmic time and
large scale structure environment in a field of 2 deg$^2$ area
\citep{cosmos-overview}. In this paper, we build mainly on the radio
(VLA-COSMOS, \citealt{schinnerer07}), X-ray (\chandra-COSMOS, or
C-COSMOS, \citealt{ccosmos-cat}), and optical spectroscopic (Z-COSMOS,
\citealt{zcosmos}) observations.  The radio data were taken at 1.4 GHz
and have a
RMS noise 7--10 $\mu$Jy (with the faintest sources
discussed in this paper having fluxes around 60$\mu$Jy), while the
X-ray data have a flux
limit of $1.9\e{-16}$ \ergscmq\ in the 0.5-—2 keV band. The area
considered here is that covered by \chandra, which is a fraction (0.9
deg$^2$) of the whole COSMOS field.  The \xmm\ observations
(XMM-COSMOS, \citealt{xmm-cosmos}) covered the whole field but with a
brighter flux limit ($1.7\e{-15}$ \ergscmq\ in the 0.5--2 keV band).
We focus on C-COSMOS here because its combination of area and limiting
flux offers the best trade off for the subject of our study.

The structure of this paper is as follows.  In
Sect.~\ref{sec:selezione} we define a sample of galaxies based on
radio and optical selection, and subsequent X-ray detection. An AGN
sample is selected with the same method to allow for comparisons.  In
Sect.~\ref{sec:optical-radio} we characterize the sample, in terms of
magnitudes, redshifts, and optical spectra.  In
Sect.~\ref{sec:characterization} we review some commonly used
X-ray-based indicators of star formation vs.\ AGN activity, and test
them on our sample of galaxies; on this basis, a refined sample is
then defined. In Sect.~\ref{sec:stacking} we investigate the average
properties of galaxies with the same radio-optical selection but
without a detection by \chandra. In Sect.~\ref{sec:misclassif}
  we discuss the number of composite and mis-classified sources. In
Sect.~\ref{sec:flussi} we consider if the COSMOS data can further
constrain the radio/X-ray correlation. Sect.~\ref{sec:lognlogs} is
devoted to an analysis of the X-ray number counts of the
radio-selected COSMOS galaxies. Finally, in
Sect.~\ref{sec:conclusions} we review our conclusions.

The cosmological parameters assumed in this paper are $H_0=70$ km s
Mpc$^{-1}$, $\Omega_\Lambda=0.7$ and $\Omega_{\mathrm M}=0.3$.

\section{Selection criteria}
\label{sec:selezione}

The catalogue of the COSMOS radio sources was published in
\citet{schinnerer07}; the objects in this catalogue were then
classified by S08, on the basis of their
photometry-based method. Only objects with redshift $\le 1.2$
were considered by S08, because the errors on the classification would be
larger beyond this threshold.
AGN can be broadly divided in two
classes: objects where the AGN dominates the entire Spectral Energy
Distribution (SED), i.e.\ mainly QSO, and objects where it does not,
such as type-II QSO, low luminosity AGN (Seyfert and LINER galaxies),
and absorption-line AGN.

 The fraction of type-I QSO in the VLA-COSMOS catalogue is
  small ($\sim 5\%$). They have higher X-ray luminosities than
  the SF and the other kinds of AGN, making the few type-I objects
  easier to separate from other classes of objects\footnote{The number
    of Chandra-detected type-I QSO is 11, out of 33 which are in the
    C-COSMOS field of view. The minimum 2-10 keV luminosity of these
    type-I QSO is $2.6\e{43}$ \ergs, therefore all type-I QSO would be
    selected as QSO/AGN (as opposed to galaxies) by the absolute
    luminosity criterion of Sect.~\ref{sec:lum42}. More in general the
    type-I and type-II classes have a different distribution of
    luminosities: the median 2--10 keV luminosities are $6\e{43}$ and
    $7\e{42}$ \ergs, respectively. Thus, we regard a simple X-ray
    luminosity argument to be sufficient to differentiate the S08
    type-I QSO from the star forming galaxy and the type-II
    populations.}; since we are mainly interested in the properties of
  galaxies, we will not discuss them further.  Hereafter with the term
  ``AGN'' we will only refer to the other kind of AGN (type-II and
  low-luminosity).  These AGN have broad-band properties similar to
  those of SF galaxies.  The main tools to disentangle SF galaxies and
  low-luminosity AGN are spectroscopic diagnostic diagrams \citep[BPT
  diagrams][]{baldwin81,vo87,kewley}, which rely on the [\ion{O}{III}
  5007]/H$\beta$ and [\ion{N}{II} 6584]/H$\alpha$ line ratios. Their
  main drawback is, however, the long telescope time needed to obtain
  good-quality spectra. Alternative methods which only build on
  photometric data can therefore be useful.  

Based on the observation of a tight correlation between rest-frame
colours of emission-line galaxies and their position in the BPT
diagram, \citet{smolcic06} used the Sloan Digital Sky Survey (SDSS)
photometry (a modified Str\"omgren system) to i) calculate rest-frame
colours, and ii) use the Principal Component Analysis (PCA) technique
to identify, among all linear combinations of colours, those which
correlate best with the position in the BPT diagram. One of these
combinations, named $P1$, was found to correlate strong enough with
the emission line properties of SF galaxies and AGN, to be used alone
for the classification.  

By applying this method to the COSMOS multi-band photometric data,
S08 produced a list of 340 `star forming' (hereafter SF)
and 601 type II/dusty/low luminosity AGN candidates.   The \chandra\ field, which is
smaller than the radio-surveyed area, contains 242 SF and 398 AGN.
Some mis-classifications are inherent in any colour- or line-based
methods, and a fraction of objects may also exhibit composite or
intermediate properties. Thus, S08 estimate that SF
samples actually may contain $\sim 20\%$ of AGN and $\sim 10\%$ of
composite objects. Conversely, AGN samples contain $\sim 5\%$ SF and
$\sim 15\%$ composite.

\begin{figure}
  \centering
  \includegraphics[width=.9\columnwidth]{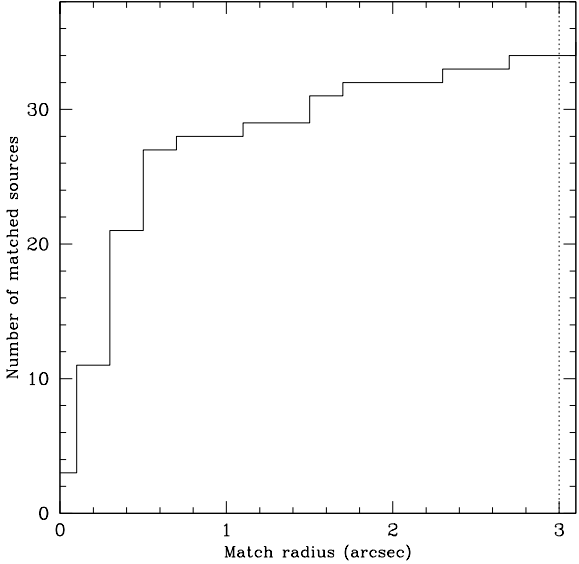}
  \caption{Number of matched sources in cross-correlating the C-COSMOS
  catalogue \citep{ccosmos-cat} with the S08 sample of
  radio-selected star forming galaxies.}
  \label{fig:crosscorr}
\end{figure}

We matched the radio positions of the SF and AGN sources
with those of the C-COSMOS
catalogue \citep{ccosmos-cat,puccetti09}. In Fig.~\ref{fig:crosscorr}
we show the number of matched SF sources for different matching radii:
the number of matches rises steeply from $0.1\arcsec$ to $0.5\arcsec$,
and flattens for larger radii. To adopt a threshold for the maximum
separation between the radio and X-ray coordinates, we considered that
in the C-COSMOS survey some areas have been observed by \chandra\ only
at large off-axis angles. For these areas, the point spread function
(PSF) is much broader than the on-axis value ($0.5\arcsec$ FWHM), and
this can also introduce errors in the determination of the source
position. The position errors reported in the C-COSMOS catalogue are
in fact larger than $1\arcsec$ for 221 sources out of 1761. Thus we
considered all matches within $3\arcsec$, and visually inspected every
match to check that the X-ray PSFs and the errors on the X-ray
positions were wide enough to justify the larger threshold. Following
this criterion, one match was excluded because the PSF in that
position was much narrower than the distance between the radio and
X-ray positions.

The samples of radio selected, optically classified, X-ray detected
sources consist thus of 33 SF ($\sim 14\%$ of the SF-classified radio
sources) and 82 AGN-type objects ($\sim 21\%$ of the AGN-classified
radio sources). The breakdown of the sources according to their
detection in the 0.5--2 keV, 2--7 keV and 0.5--7 keV bands is shown in
Table~\ref{tab:num_src}.  Note the presence of 10 SF candidates
lacking a detection in the soft band: this hints for the presence of
AGN-type objects in the SF sample, which will be discussed in detail
in the next sections.

\begin{table}
  \centering
  \begin{tabular}{lrr}
    Band        &SF    &AGN  \\ \hline
    F+S+H       &10    &46        \\
    F+S         &11    &16        \\
    F+H         &10    &18        \\
    F           &---   &1         \\
    S           &2     &1         \\
    H           &---   &---       \\ \hline
  \end{tabular}
  \caption{Number of X-ray detected radio sources, according to their
    optical classification and X-ray band of detection (F: full,
    0.5--7 keV; S: soft, 0.5--2 keV; H: hard, 2-7 keV).}
  \label{tab:num_src}
\end{table}

\section{Optical and radio properties}
\label{sec:optical-radio}

\begin{figure*}
  \centering
   \includegraphics[width=.33\textwidth,angle=-90]{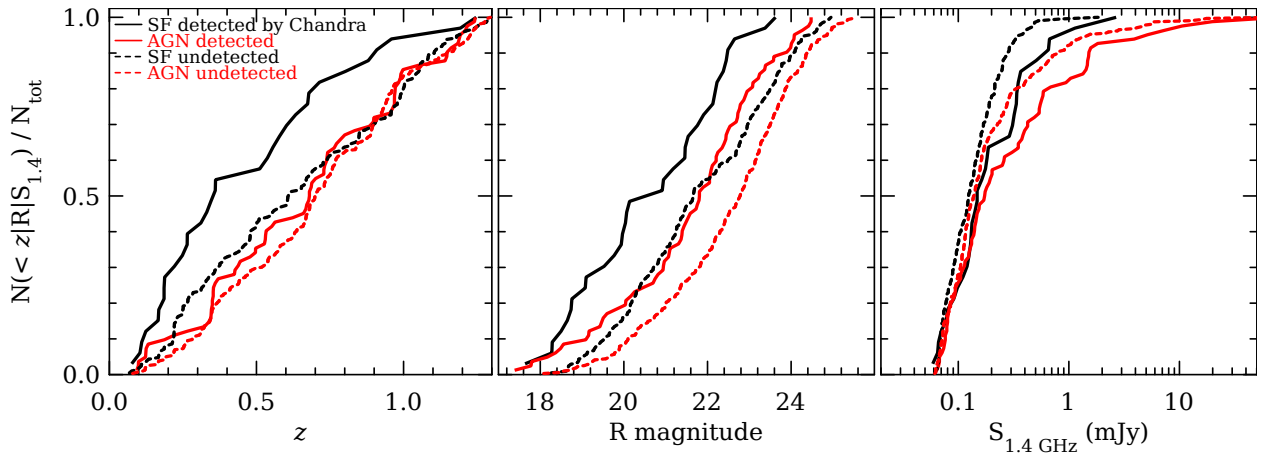}

  \caption{Relative cumulative distributions of redshift, R magnitude
    and radio flux in the \chandra-detected and undetected, SF and AGN
    galaxies subsamples.}
  \label{fig:samplehistograms}
\end{figure*}

The samples of X-ray detected vs.\ undetected sources exhibit
different properties, as shown both in the following tests and in the
cumulative distributions in Fig.~\ref{fig:samplehistograms}. For the
SF sample we find that:
\begin{itemize}
\item 
detected sources have brighter (observer-frame) R magnitudes than
undetected ones, at a confidence level of 99.96\% (according to a
Wilcoxon-Mann-Whitney test; the median magnitudes are $R=20.53$ and
21.66 for the detected and undetected, respectively);

\item 
detected sources have brighter radio fluxes than undetected, at a
confidence level of 98.5\% (median fluxes: $\sradio =0.148$ and 0.124 mJy,
respectively);

\item 
detected sources have lower redshifts than undetected, at a
confidence level of 99.8\% (median redshifts: $z=0.36$ and 0.61,
respectively);

\item 
detected sources have lower radio luminosities than undetected,
at a confidence level of 97.8\% (median luminosities: $S=8.7\e{29}$
and $5.0\e{30}$ \ergsHz, respectively).
\end{itemize}
This is in line with the expectation of the detected sources being
closer to us, and the undetected ones probing a larger volume where more
luminous yet less common objects can be found.

The X-ray detected AGN sources also have brighter R
magnitudes and radio fluxes than the undetected ones, but do not show
any significant difference in the redshift distribution. The
behaviour of the radio luminosity is reversed: the undetected AGN
have lower radio luminosities than the detected ones.

\subsection*{Optical spectra of X-ray detected sources}
\label{sec:spectra}

Optical spectra are available for most of the sources with an X-ray
detection from several spectroscopic campaigns: the ZCOSMOS project
\citep{zcosmos,zcosmos10k}, Magellan/IMACS surveys, the Sloan Digital
Sky Survey (SDSS), \citep{trump07,trump09} and deeper
observations\footnote{PIs: Capak, Kartalpepe, Salvato, Sanders,
  Scoville.}  with Keck/DEIMOS and VIMOS/VLT.  A simple classification
based on diagnostic diagrams \citep{bongiorno10}, further checked by
visual inspection, has been used to the determine the optical
classifications \citep{civano12}.  Since the signal/noise ratio varies
a lot in the sample, some objects have noisy spectra which can only
be classified tentatively.

In the SF sample, 21 objects (out of 33) are classified as emission
line galaxies; 9 are classified as AGN, and 3 have no spectral
information or have spectra with low signal/noise ratios.

In the AGN sample, 25 objects (out of 82) are classified as AGN; 30 as
emission line galaxies; 7 as absorption line galaxies, and 20 have no
spectral information or spectra with low signal/noise ratio.

This partial overlap in the optical classification between the two
catalogues is expected (see also Sect.~\ref{sec:selezione}), because 
{\em i)} the two phenomena of accretion and star formation are often
present together in the same object,
{\em ii)} the overlap of the areas covered by different populations
(SF and AGN) in the diagnostic diagrams used by S08,
{\em iii)} low-luminosity, narrow-line AGN and actively star forming
galaxies can be difficult to distinguish in noisy spectra.
The last point is particularly true at $z\gtrsim 0.4$, where the
H$\alpha$ line is not sampled by optical spectra and, therefore, the
standard BPT diagram ([O \textsc{III}]/H$\beta$ vs.\ [H
\textsc{II}]/H$\alpha$; \citealt{baldwin81}) cannot be used in the
optical spectral classification.
A comparison of the observed fraction of mis-classifications
and composite objects with the expectations will be presented in
Sect.~\ref{sec:misclassif}.

In the following Section, we will try and characterize
further the two samples on the basis of the X-ray properties of
the sources, with the aim of improving the classification.

\section{X-ray characterization of the selected sources}
\label{sec:characterization}
X-ray spectra of SF galaxies are rather complex, as they include
emission from hot
gas, supernova remnants (thermal spectra) and X-ray binaries
(non-thermal, power-law spectrum), with the thermal components being
usually softer than the non-thermal ones. A detailed description of
the expected spectra of the different components may be found in
\citet{pr02}. The relative importance of the spectral components may
vary; however, in most cases the average flux ratio between the
0.5--2.0 keV and the 2.0--10 keV bands is the same that would be
obtained if the spectrum was a power-law spectrum with spectral index
$\Gamma=2.1$ and negligible absorption (RCS03; \citealt{lehmer08}).  This
does not imply the lack of X-ray absorption of X-rays in SF galaxies:
M82 and NGC3256 are notable examples (RCS03; \citealt{m82centok}). However,
the spectral analysis of 23 SF galaxies in RCS03 did not
find heavy absorption to be a general property of that sample.

The fluxes of candidate SF objects used in this paper have been
recomputed from the counts by assuming the $\Gamma=2.1$ spectrum,
instead of the $\Gamma=1.4$ used in \citet{ccosmos-cat}. Conversely,
for AGN we have used the latter, harder spectrum. In the following, we
review a few common indicators of SF vs.\ AGN activity, and apply them
to the SF sample for further screening.

\subsection{Hardness ratio}
\label{sec:HR}

\begin{figure*}
  \centering
  \includegraphics[width=.995\columnwidth]{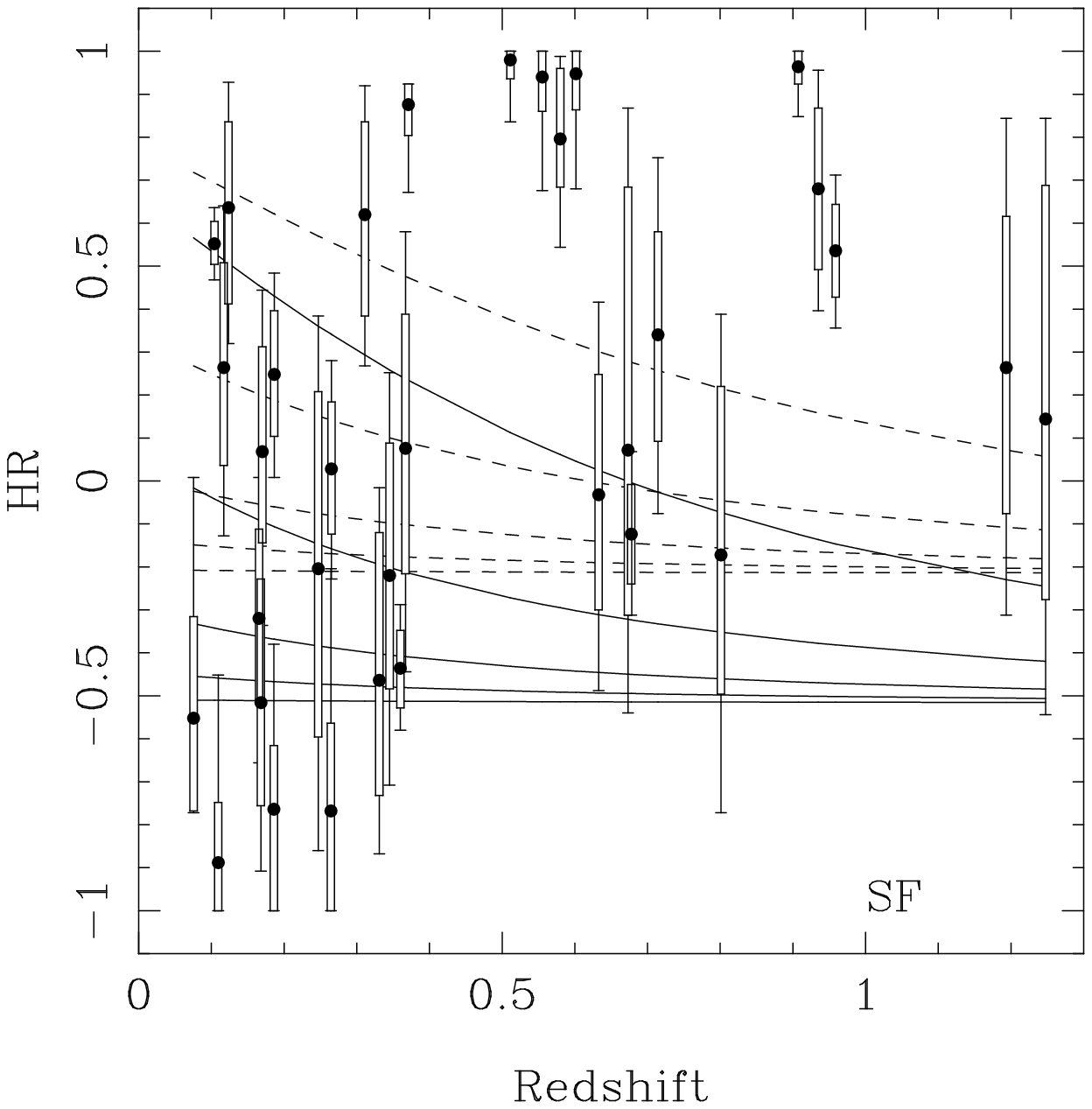}
  \hfill
  \includegraphics[width=.995\columnwidth]{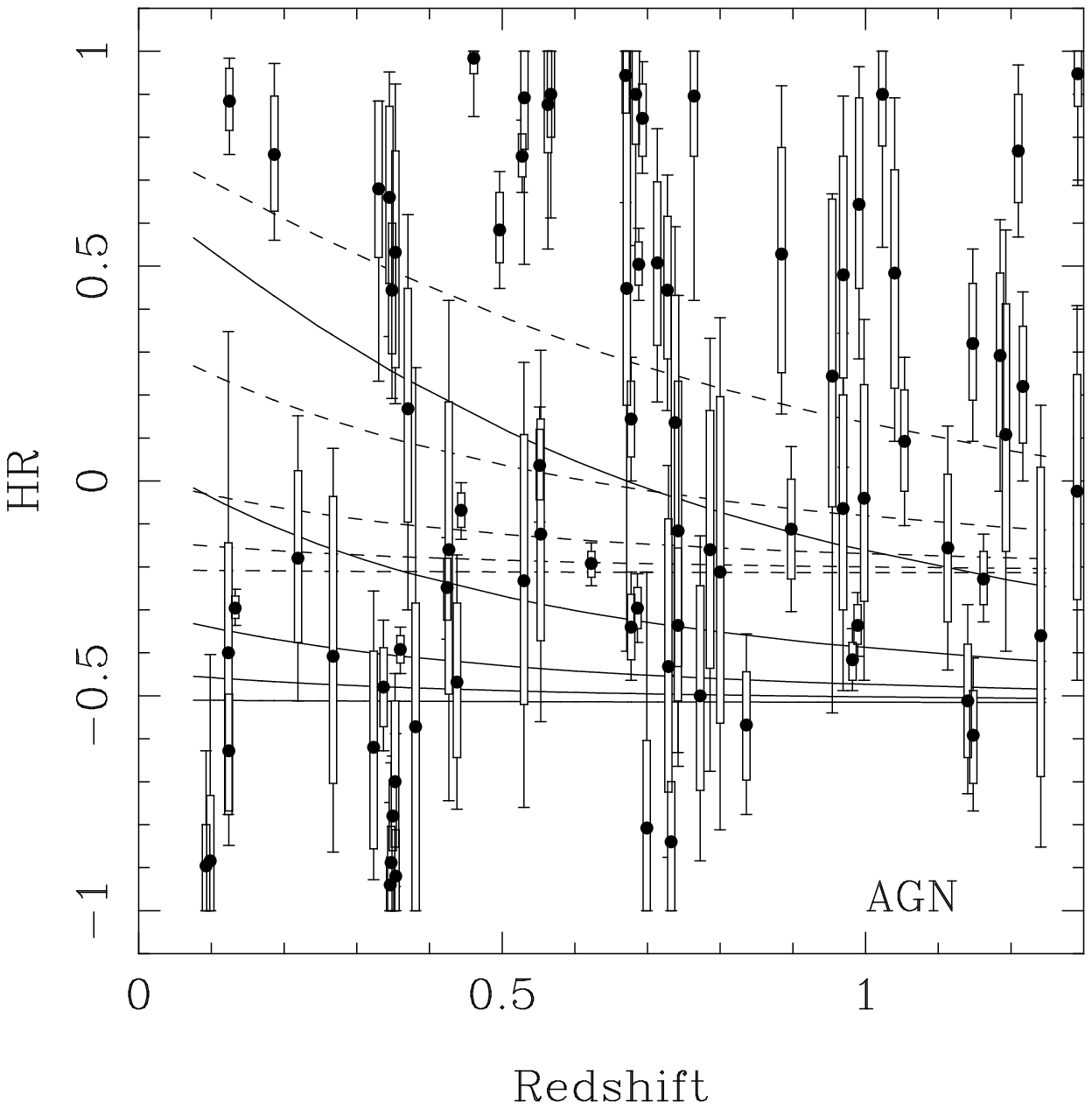}
  \caption{Hardness ratios for the SF (left) and AGN (right) X-ray
    detected sources. The data
    points, boxes and whiskers show the median and the 68\% and 90\%
    credible intervals for the HR, respectively. The superimposed
    lines show the HR referring to model spectra: absorbed power-laws
    with two different slopes ($\Gamma=2.1$, solid lines, and
    $\Gamma=1.4$, dotted lines) and five column densities ($N_{\rm
      H}=10^{20}$, $10^{21}$, $10^{21.5}$, $10^{22}$ and $10^{22.5}$ cm$^{-2}$ from
    bottom to top).}
  \label{fig:behr}
\end{figure*}

The hardness ratio (HR) is a simple measure of the spectral shape,
defined as $HR=(H-S)/(H+S)$, where $H$ and $S$ are the net (i.e.\
background-subtracted) counts in the hard and soft bands respectively.
It is most useful when the sources are too faint for a proper spectral
analysis. The error on the HR is usually calculated by taking the
uncertainties on source and background counts according to the
\citet{gehrels86} approximation, and applying the error propagation
for Gaussian distributions. However, it has been shown that for faint
sources this approach largely overestimates the errors on the HR
\citep{behr}. To obtain a more realistic estimate of the HR
uncertainties, we use the Bayesian approach described in
\citet{behr}. The source and local background counts have been
extracted from the \chandra\ event files in soft (0.5--2 keV) and
hard (2--7 keV) observer-frame energy bands.
For the
sources in the SF and AGN samples we show in Fig.~\ref{fig:behr} the median
value of the HR posterior distributions, and the 68\% and 90\% highest
probability density (HPD) intervals of the posterior distributions,
vs.\ the redshifts. The HR corresponding to eight model spectra
(absorbed power-laws) are also superimposed.
These model HR have
been calculated with XSPEC, using average effective areas (see
Sect.~\ref{sec:stacking}) and assuming no background.

About half of the SF sources detected in C-COSMOS have a HR which is
consistent with very flat or absorbed spectra ($N_{\rm H}\gtrsim
10^{21.5}$ cm$^{-2}$, or equivalently $\Gamma\lesssim 1.2$ without
absorption). Conversely, about half of the AGN sources have a
HR consistent with spectra steeper or less absorbed than the above thresholds.

The average hardness ratio of galaxies is the one given by a
non-absorbed $\Gamma=2.1$ spectrum (RCS03). Harder spectra are
sometimes indeed found in galaxies (e.g., see the slopes for single
power-law fits in \citealt{dahlem98}) where they can result from
bright X-ray binaries \citep{pr02}.  However, \citet{swartz04} found
that the average slopes of X-ray binaries in nearby galaxies are
$\Gamma=1.88\pm 0.06$ ($1.97\pm 0.11$) for binaries with luminosity
larger (smaller) than $10^{39}$ \ergs, respectively. Only 7\% (8\%)
of the binaries studied by \citet{swartz04} have slopes $\Gamma\le
1.4$.

The large number of hard objects in the SF sample is probably due to a
selection effect. Because of the C-COSMOS flux limit, the
\chandra-detected SF sample only contains the brightest 14\% of all
the SF objects in the field of view.  In addition, 30\% of the
S08 SF sources are expected to be composite or
misclassified, as is inherent in any selection method based on
diagnostic diagrams. Since AGN are on average brighter than galaxies,
the composite/misclassified objects should mingle with the brightest galaxies,
hence with the \chandra-detected sample rather than with the
\chandra-undetected one.

\subsection{X-ray/optical flux ratio}
\label{sec:X/O}

A fast and widely used, yet coarse method to classify X-ray objects is
to look at their X-ray/optical flux ratio $X/O$, defined as
\begin{equation}
  \label{eq:rapp_xottico}
  X/O = \Log \left( F_{\rm X} \right) + 0.4 R + 5.71
\end{equation}
where $F_{\rm X}$ is the 0.5--2 keV flux, and $R$ is the optical
apparent magnitude in the R filter. 
On average, AGN have higher $X/O$ values than SF
galaxies. Given the intrinsic dispersion in the $X/O$ values for
both AGN and SF galaxies, no single $X/O$ value can be used to
unambiguously separate AGN from SF galaxies. However, it has been
shown \citep{schmidt98,bauer04,alexander02} that the value $X/O=-1$
can be
taken as a rough boundary between objects powered by star formation
($X/O<-1$) and by nuclear activity ($X/O>-1$). In the SF sample, 22
objects have $X/O<-1$ (2/3 of the total sample), while 11 have $X/O>-1$. Out of this latter 11
objects, which this criterion would classify as AGN, 10 have a HR
whose 68\% HPD interval is consistent with a column density $N_{\rm
  H}\gtrsim 10^{21.5}$ cm$^{-2}$.

For comparison, the AGN sample has only 1/3 of the objects with
$X/O<-1$ (27 objects out of 82).

\subsection{X-ray luminosity}
\label{sec:lum42}

An X-ray luminosity of $10^{42}$ erg/s is also often used as another
rough boundary between SF galaxies and AGN, with the band in which the
luminosity is measured varying among different authors.  While this
criterion is, on its own, so unrefined that it ignores the existence
of low luminosity AGN, it may still be of use when considered along with
other criteria. In the SF sample, 17 objects have a 2--10 keV luminosity
greater than this limit. However, it is expected that the X-ray
luminosity of SF galaxies evolves with the redshift; RCS05
found that a pure luminosity evolution of the form $L_{\rm X} \propto
(1+z)^{2.7}$ is a good description of the available data \citep[see
also][]{norman04}. Thus one should consider as AGN candidates only the
objects with $L_{\rm X}> 10^{42}(1+z)^{2.7}$ \ergs: 12 objects in the SF
sample satisfy this criterion. All these 12 objects have a HR
compatible with a column density larger than $10^{21.5}$ cm$^{-2}$.

\newcommand{\soft}{(soft)}
\newcommand{\hard}{(hard)}

\addtocounter{table}{1} 
\addtocounter{table}{1}

\subsection{Off-nuclear sources}
\label{sec:offnuclear}

If the X-ray position is not coincident with the galaxy centre, but is
still within the area covered by the galaxy in the optical band, then
the X-ray emission is probably due to an off-nuclear X-ray
binary. Thus any contribution from nuclear accretion is unlikely to be
significant.
An off-nuclear flag is present in the catalogue of optical
identifications (\citealt{civano12}; see also \citealt{mainieri10}).
The only SF source classified as off-nuclear is
CXOC\,J100058.6+021139. No source from the AGN sample is classified as
off-nuclear. While in principle this is an important selection
criterion, in practice it applies to only one object in our samples,
and therefore we will not consider it any further.

\subsection{Classification and catalogue of sources}
\label{sec:cat}

From the considerations made above, it seems that only
about half of the objects in the SF sample have properties compatible
with SF-powered X-ray emission.  This hints to the presence of several
objects in the sample which show intermediate, rather than SF
or AGN properties. Some refinement of the S08 SF
selection criteria seems thus possible by inspecting the X-ray
properties of the sources.

We consider the following conditions as indicators of SF origin of the
X-ray luminosity:
\begin{itemize}
\item $L_{2-10}\le 10^{42}(1+z)^{2.7}$ \ergs, where $L_{2-10}$ is the hard X-ray
  (2.0-10 keV) luminosity;
\item HR lower (softer) than what expected for an absorbed power-law with
  $\Gamma=1.4$ and $N_\mathrm{H}= 10^{22}$ cm$^{-2}$;
\item $X/O\le -1$;
\item classification as galaxy from optical spectroscopy.
\end{itemize}

For the purposes of classification,
we only consider optical spectra with clear AGN-like emission line
ratios as non-matching. This avoids that low signal/noise spectra can
influence the classification.

Among many possibilities to combine to above criteria, we use the following
method: the number of matched criteria is counted, and
an object is classified accordingly. An object is assigned to class 1,
if it fulfils all the conditions; to class 2, if it fulfils all
conditions but one; to class 3, if there are at least two conditions
not satisfied.  Objects for which one condition can not be checked
(e.g., a missing optical spectrum), are classified as
if that condition had been matched. The idea is that an object status
is affected only by conditions which have been checked and not
matched.

Thus we recognize two samples of SF galaxies: one more conservative
(class 1 objects), and another less conservative (objects in classes 1
or 2). Class 3 objects probably do not have the
majority of their X-ray emission powered by star formation related
processes.  There are 8 class-1 objects; 8 class-2; and 17 class-3
objects in the SF sample.

If the same selection is applied to the AGN sample, we find 14
class-1; 6 class-2; and 62 class-3.

The classifications for both the SF and AGN sample are reported in
Tables~\ref{tab:cat} and \ref{tab:catAGN}, along with
the other parameters of interest: X-ray fluxes, luminosities, medians of the
HR posterior probability distributions, $X/O$ ratios, X-ray/radio
ratios (see Sect.~\ref{sec:flussi}), classification from optical
spectroscopy.

\section{Average properties of undetected objects}
\label{sec:stacking}

\begin{table*}
  \caption{Average properties of radio-selected SF-candidates
    without an X-ray detection in C-COSMOS. Radio fluxes in mJy;
    X-ray fluxes in $10^{-18}$ \ergscmq; X-ray luminosities (rest
    frame) in $10^{40}$ \ergs. The counts are reported with their
    68.3\% error intervals. The fluxes have been calculated by fitting
    the stacked spectra with an imposed $\Gamma=2.1$ (this slope has
    the characteristic of making the 0.5--2 and 2--10 keV fluxes
    approximately the same; considering the rounding of
    non-significant digits, this explains why the fluxes are 
    the same in the two bands). All values for the \textit{soft band}
    are relative to the 0.5--2 keV interval,
    while for the \textit{hard band} counts are in the 2--7 keV,
    and fluxes and luminosities are in the 2--10 keV
    bands. The allowed spectral slope ($\Gamma$) range is
    given in the last column.}
  \label{tab:stacking}
  \centering
  \begin{tabular}{lrrrlrrrrrrr}
    \hline
    Selection &No.\ of         &Avg.\ radio &Avg.     &Band   &Exposure &Net   &X-ray&Lumi- &$1-p_{\rm detect}$ &$1-p_{\rm cts}$ &$\Gamma$ \\
              &candidates      &flux (mJy)  &redshift &       &(Ms)     &counts&Flux &nosity&                 &              &\\ 
    \hline
    $\sradio\le 0.20$
                    & 156      &0.113       &0.67     &soft &16.9     &$176\pm20$ &4.8  &0.94 &$99.99\%$ &$99.99\%$          &[1.5--\\
       "            & "        & "          & "       &hard &17.7     &$105\pm25$ &4.8  &0.94 &$99.98\%$ &$>99.99\%$         &2.5]\\
    $0.20<\sradio\le 0.63\!\!\!\!\!\!\!\!$ 
                    & 43       &0.292       &0.58     &soft &4.3      &$80\pm11$       &8.2  &1.1 &$99.96\%$ &$99.92\%$      &[1.7--\\
       "            & "        & "          & "       &hard &4.5      &$16^{+4.7}_{-10}$ &8.2  &1.1 &---       &---            &3.8]\\
  \hline
  \end{tabular}
\end{table*}

The method of stacking analysis allows to determine the average
properties of objects which are not individually detected; it can be
briefly described as follows.  Candidate objects for stacking are
selected from the list of SF sources in S08 which are
not detected in C-COSMOS, with the additional criterion that no
detected C-COSMOS source should be present within 7 arcsec from the
position of the candidate. The reason is to avoid contamination from
X-ray brighter sources. This does not introduce any bias in the
selection of sources, because very few sources are excluded in this
way (only 6 out of 209).

Because of the radio--X-ray correlation, it is expected that the
average X-ray properties are dominated by the brightest radio sources. Thus
it may be advisable to include in the stack only sources with a narrow
spread in their radio flux, to avoid biases due to the brightest
sources.  We split the sample on the basis of the radio flux, dividing
the candidates in two lists as follows:
\begin{enumerate}
\item 156 SF sources with $\sradio\le 0.20 $mJy;
\item 43 SF sources with $0.20< \sradio\le 0.63$ mJy.
\end{enumerate}
Each sub-sample is 0.5 dex wide in flux; one starts from
the lowest radio flux, while the other one follows continuously. 

For each list of candidates, postage-stamp size images measuring
$20\times 20$ pixels (each one $0.491\arcsec$ wide) around the
position of every candidate, are extracted and summed; the latter sum
is hereafter called ``stacked image''. If most candidates contribute
some X-ray photons, then a ``stacked source'' appears on top of the
background in the centre of the stacked image (similar results can be
obtained using the software described in \citealt{miyaji_stack}). The
{\tt wavdetect} tool is then used to determine the net counts of the
stacked source. This analysis was done for both the 0.5--2.0 keV and
2.0--7.0 keV bands.

The stacked source was successfully detected by {\tt wavdetect} for
the low-radio flux sub-sample in both bands, and for the high-radio
flux in the soft band only. This is in line with the expectation that
a lower number of counts should be present in the high- than
in the low-radio-flux subsample: given the average fluxes and the number
of positions (Table~\ref{tab:stacking}), $\sim 40\%$ more counts are
expected in the latter than in the former. 

The significance of the detection of a stacked source is best
estimated with simulations: we draw as many random positions as the
number of sources in the list, reproducing the same spatial
distribution of the VLA-COSMOS sources, and build a stacked image from
these positions. Total counts ($c_{\rm sim}$) within a radius of
3.5\arcsec\ from the centre are extracted; the {\tt wavdetect} software
is run on the stacked image; this cycle is repeated 10000 times for
each sub-sample and each band. We then define the following $p$-values:
\begin{itemize}
\item $p_{\rm detect}$ as the fraction of times that the {\tt wavdetect}
  software finds a source within 1.1 arcsec of the centre of the stacked image;
\item $p_{\rm cts}$ as the fraction of times that $c_{\rm sim} \ge
  c_{\rm stack}$, where $c_{\rm stack}$ are the total counts of the
  `real' stacked source.
\end{itemize}
We identify the $p$-values as two estimates of the probability that
the stacked source was actually a background fluctuation.
The $p$-values are shown in
Table~\ref{tab:stacking} (actually, $1-p$ is shown, i.e.\ the
probability that the source is not a fluctuation).

Using the source and background regions defined above, and a power-law
average spectrum with $\Gamma=2.1$, we extracted the net counts and
derived the fluxes and luminosities shown in Table~\ref{tab:stacking}.

Stacked X-ray spectra have also been extracted for the two subsamples, using
CIAO 4.0.  Background spectra have been extracted around the source
positions, by taking 4 circular background regions for each source,
each background region having the same radius of the source region, and being
placed $10\arcsec$ east, north, west, or south of the source. This
ensures that the background is the most accurate, given the actual sky
positions of the sources. Then, we removed background positions which
fell within $7\arcsec$ from any \chandra-detected source.
Response matrices have been calculated considering the stacked source
like it was an extended source consisting of many small pieces
scattered around the detector area, weighted by the photons actually
present in each position.

\begin{figure}
  \centering
 \includegraphics[width=.49\textwidth]{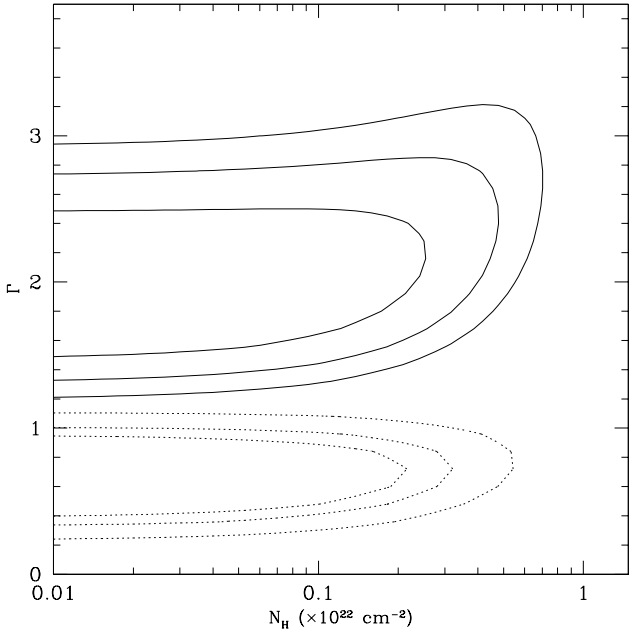}
  \caption{Confidence contours for relevant parameters of simple
    models of stacked spectra of SF sources and (for comparison) AGN sources
    not detected by \chandra\ 
    (sources with radio flux $F<0.20$
    mJy).
    The contours are shown at levels of
    $\Delta \mathrm{C}$: +2.92, +5.63, +8.34 above minima.
    Solid lines: SF sources; dotted lines: AGN.}
  \label{fig:stackedspectra}
\end{figure}

The stacked spectra were fitted with a model which is the weighted sum
of many absorbed power-laws; the number of power-law components is the
number of sources in the stack. Each absorbed power-law is redshifted
to the $z$ of the corresponding source. The slope and absorption are
free parameters, but are assumed to be the same for all sources. The
weights are proportional to the radio fluxes.
Finally,
observer-frame absorption due to the Galaxy is added. 
This method allows to fully account for the redshift
distribution of the sources. 
We used XSPEC version 11.3.2 for this analysis.

The confidence contours for the parameters are shown in
Fig.~\ref{fig:stackedspectra} (solid curves) for the bin with sources
with $\sradio\le 0.2$ mJy. The spectrum is consistent with moderately steep
photon indices ($1.5\lesssim \Gamma \lesssim 2.5$). This behaviour is
expected for star forming galaxies (see
Sect.~\ref{sec:characterization}).  The bin with sources with
$0.2<\sradio\le 0.63$ mJy (not shown) has a lower number of X-ray photons in
the spectrum and thus a wider range of slopes allowed, yet the
spectrum is still consistent with the other bin.

\begin{figure}
  \centering
   \includegraphics[width=\columnwidth]{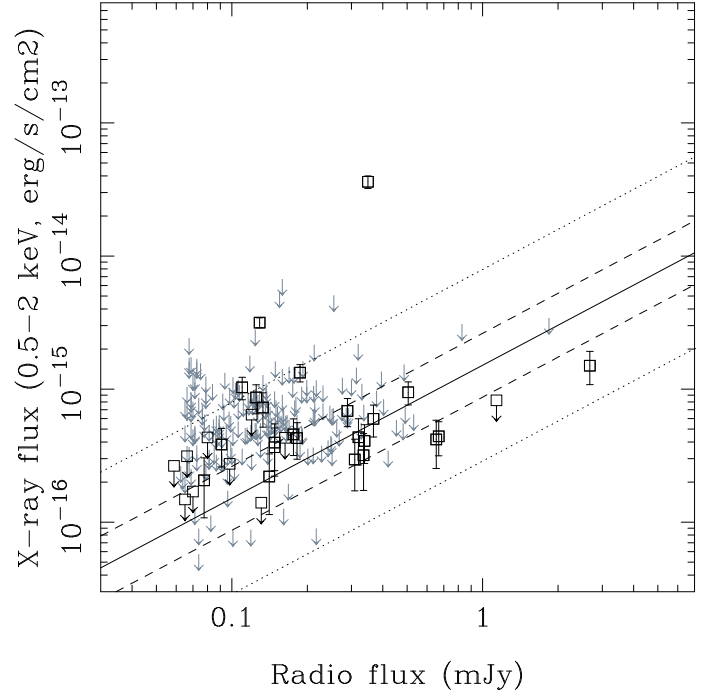}
  \caption{Radio vs.\ X-ray (0.5--2 keV) fluxes for the VLA-COSMOS SF sources detected
    in C-COSMOS. 
    Detected sources are marked with squares
    and error bars. The black squares with an attached arrow
      mark the sources which are not detected in the band
    to which the panel refers, but are detected in any other of the
    C-COSMOS bands. Conversely, the grey upper limits show the
    sources without a detection in any X-ray band, hence without an
    entry in the C-COSMOS catalogue.
    The solid line shows the
    RCS03 relationship, K-corrected to the average redshift of the
    detected sources ($z=0.46$), while the dashed lines show the 1$\times$ and
    3$\times$ standard deviation of the relationship.
  }
  \label{fig:radiox}
\end{figure}

For comparison, in Fig.~\ref{fig:stackedspectra} (dotted curves) we
show also the confidence contours for stacked spectra of 207 AGN,
selected from the S08 sample in the same radio flux
intervals of the SF galaxies, and also not detected by \chandra. The
average redshifts of these AGN are not significantly different from
the SF ones.  The AGN X-ray spectra are flatter; this is probably the
result of absorption and of redshift (summing absorbed spectra of
sources at different redshift leads to flat or inverted spectra, like
in the case of the cosmic X-ray background). While a single power-law
model is probably simplistic, a more detailed modeling would be beyond
the scope of this paper.
The average X-ray fluxes for the AGN, calculated with the best-fit
slopes, are $7.2\e{-18}$ ($4.8\e{-17}$) \ergscmq\ for the bin with
$\sradio\le 0.2$ mJy and $3.0\e{-18}$ ($1.0\e{-17}$) \ergscmq\ for the
bin with $0.2<\sradio\le 0.63$ mJy in the 0.5--2 (2--10) keV band. 
The allowed ranges for the AGN spectral slopes are [0.4--0.9] and
[0.6--2.2] for the $\sradio\le 0.2$ mJy and $0.2<\sradio\le 0.63$ mJy
bins, respectively.

\section{Mis-classified and composite objects}
\label{sec:misclassif}

The fractions of X-ray detected objects whose classification
based on optical spectral line ratios
(where available) is different from that based on the synthetic $P1$ colour in S08 are
$9/33\sim 27\%$ (SF sample) and $37/82\sim 45\%$ (AGN sample).  These
numbers are of the same magnitude of those quoted in
Sects.~\ref{sec:HR},\ref{sec:X/O},~\ref{sec:lum42}, though it is
important to stress that different criteria yield different
mis-classified and composite (hereafter MCC) objects. 

These fractions should be compared with the number
of class 3 objects in the SF sample ($17/33\sim 51\%$) and of classes
1-2 in the AGN sample ($20/82\sim 24\%$).

However, these fractions do not take into account the large number
of X-ray undetected objects and should probably be
considered as upper limits to the true fractions of MCC objects. In
fact, the stark difference found between the
average X-ray spectra of undetected SF and AGN sources
(Sect.\ref{sec:stacking}) would rather suggest much lower fractions of
MCC. A lower limit to the true fractions of MCC can be derived by assuming
that the MCC are only present among the X-ray
detected sources. In this case, the fractions would be $17/242\sim 7\%$ ($20/398\sim
5\%$) for the class 3 (classes 1-2) in the SF (AGN) sample. The
rationale for this assumption would be, for the SF sample, that
galaxies with an AGN contribution are on average brighter in X-rays than the
galaxies without and thus are more likely to be X-ray detected. For
the AGN sample, it would be that intense star formation could cause
X-ray emission at a level similar to that of a low-luminosity or absorbed active nucleus.

The fractions reported
by S08 (30\% of MCC in the SF and 20\% in the AGN
sample; see Sect.~\ref{sec:selezione}) are intermediate between our
upper and lower limits' estimates, and
therefore we regard them as in agreement with our findings. In the
following, we use the method described in Sect.~\ref{sec:cat}
to identify the MCC
candidates. The same method cannot be applied to X-ray undetected
objects, and therefore the optical colour-based classification is used for
them in the remaining of this paper.

\section{The X-ray/radio flux ratio}
\label{sec:flussi}

Correlations between X-ray and radio luminosities of star forming
galaxies, and between the X-ray and far infrared (FIR) ones, are well
established for the local universe and have been tested for objects up
to $z\sim 1$ (\citealt{bauer02}; RCS03;
\citealt{gilfanov04a,pr07}). Both the radio and FIR luminosity are
tracers of the star formation rate (SFR). These correlations are
linear, and imply that in absence of any contribution from an AGN, the
X-ray emission is powered by star-formation related
processes. High-Mass X-ray Binaries (HMXB) seem to have the same
luminosity function in all galaxies, only normalised according to the
actual SFR. Thus, if HMXB are the dominant contributors to the X-ray
emission, the X-ray luminosity is a tracer of the SFR
\citep{grimm02,gilfanov04b,pr07}. The other possibly dominant
contributors to the X-ray emission are the Low-Mass X-ray Binaries
(LMXB), whose number scales with the galaxy stellar mass. The actual
balance of the two populations thus depends on the ratio between SFR
and mass, thus it is possible that there is some evolution of the
correlation due to the mass build-up by the galaxies with the cosmic
time.

The VLA- and C-COSMOS surveys contain a sizable number of objects at
medium-deep redshifts on which the correlation might be
tested. However, many objects have radio fluxes close to the
VLA-COSMOS flux limit. This is clearly visible in
Fig.~\ref{fig:radiox}, where we show the radio and X-ray fluxes and
upper limits for all the SF galaxies in the C-COSMOS field. Since we
account for the X-ray upper-limits, the main source of potential bias
is the radio flux limit; the results may therefore be biased towards
radio-brighter-than-X-rays objects. To partially mitigate this effect,
we split the sample in two redshift bins, defined as follows.

The knee of the radio LF of galaxies is found at a luminosity
$L_{\mathrm k}\sim 1.5\e{29}$ \ergsHz\ at 1.4 GHz (see, e.g., Fig.2 in
RCS05). We define the first bin as the redshift interval in which all
objects with luminosity $\ge \frac{1}{2} L_{\mathrm k}$ can be
observed. This corresponds to $z\le 0.2$. In other words, this
redshift threshold guarantees that the luminosities around $L_{\mathrm
  k}$ are included in the sampled volume. While this bin does not
contain a strictly volume-limited subsample, here the selection bias
should be mitigated as much as possible, and still the bin includes a
sizable number of objects. 

The second bin contains the remaining objects, i.e.\ those with
$z>0.2$.  Finally, we refer to calculations made on all objects
as a third bin named ``any-$z$''.

For most of the objects, the 0.5--2 keV flux limit is a few times
larger than what expected from the radio flux limits and the
radio/X-ray correlation. Many objects therefore only have X-ray upper
limits, which need to be properly accounted for.  The 2--10 keV limit
is about one order of magnitude larger and thus this band is not
considered in this Section.

The hypothesis we would like to test is if the COSMOS data are
consistent with the extrapolation of the RCS03 correlation, or if they
require substantially different parameters. This kind of question is
usually what Bayesian methods are most suited to answer.
The ratio
\begin{equation}
  \label{eq:q}
  q = \Log ( F_{0.5-2 \mathrm{keV}} / \sradio )
\end{equation}
may be defined in the same way of the analogous $q$ parameter often
employed in testing the radio/FIR correlation of spiral galaxies
\citep{cond92}.  Non-detections in X-rays therefore lead to upper
limits for the $q$'s and can be properly accounted for. 

To include the K-correction in the $q$'s, we consider the
individual redshifts of the sources and assume average
power-law spectra with average slopes. 
Using X-rays (0.5--2.0 keV luminosity less than $10^{42}$
\ergs) to remove bright AGN from the VLA-CDFS survey
\citep{vla-cdfs-cat,vla-cdfs-tozzi09}, a radio spectral energy index
$\alpha\sim 0.69$ can be assumed for the galaxies.

\begin{figure}
  \centering
  \includegraphics[height=\columnwidth,angle=-90]{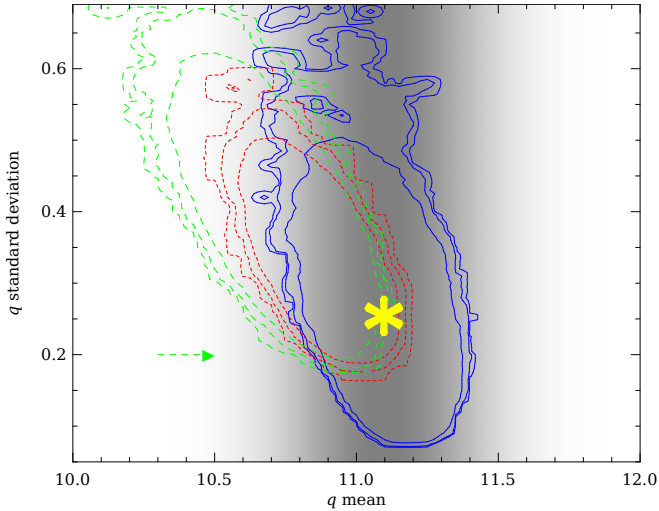}
  \caption{ 
    Credible contours for the logarithmic X-ray/radio
    ratio $q$. The Bayes formula was used to derive the
    posterior probability for $q$, shown as contours (levels
    relative to 50\%, 68.3\% and 90\%). The blue continuous and the
    green long-dashed contours refer to SF objects with $z\le 0.2$
    and $z>0.2$, respectively. 
    The red
    short-dashed contours refer to SF objects at any redshift (i.e.,
    the sum of the two bins). The green long-dashed arrow shows the
    correction that should be made to $q$ for the $z>0.2$ and any-$z$
    bins to account for the bias due to radio selection.
    The grey-scale background shows the
    prior (a darker grey means a higher probability density). The
    yellow asterisk shows the $q_0$ value from RCS03.  }
  \label{fig:qbayes}
\end{figure}

The radio/X-ray correlation is described by its slope (here assumed unity),
its mean $\bar q$, and its standard deviation $\sigma$.  The two
latter parameters are the subject of the present statistical analysis,
and need prior probability distributions, which we take as follows:
\begin{itemize}
\item $\bar q_\mathrm{prior}$ is assumed to follow a normal
  distribution with mean $\bar q_0$ and standard deviation
  $\sigma_{0}$ taken equal to the values found by RCS03 in the local
  universe ($\bar
  q_0=11.10$ and $\sigma_{0}=0.24$);
\item the standard deviation $\sigma_{\mathrm{prior}}$ is
  assumed to be uniformly distributed.
\end{itemize}
In Fig.~\ref{fig:qbayes}, the prior distribution is represented as the
grey shade.

The Bayesian posterior probabilities were calculated with the
Montecarlo method \citep{meekerescobar}; the credible
contours\footnote{It may be worth reminding that in the Bayesian
  framework one speaks of \textit{credible} contours and intervals,
  leaving the word \textit{confidence} for frequentist statistics in
  order to avoid confusion.} for the mean and standard deviations of
$q$ are also shown in Fig.~\ref{fig:qbayes}, along with the RCS03
estimate.

The posterior distribution for $q$ and $\sigma$ is consistent with
the RCS03 values, within the 68.3\% HPD area for the $z\le 0.2$ and
any-$z$ bins, and within the 90\% Highest Posterior Density (HPD) area
for the $z>0.2$ bin.  However, the centres of the posterior
distributions of the $z>0.2$ and the any-$z$ bins appear to be shifted
to lower values of $q$.

The $z>0.2$ and the any-$z$ bins allow a larger $\sigma$
than RCS03 (by a factor $\lesssim 2$), while the $z\le 0.2$ bin also
allows smaller $\sigma$. The reasons for the larger dispersion may
include the large number of upper limits, uncertainties on the
K-correction, and a residual contamination by AGN of the objects with
X-ray upper limits.

One possible explanation of a smaller $q$ at high redshift
is that the radio luminosity is evolving in redshift at a faster pace
than the X-ray one. The possibilities for an increased efficiency of
the radio emission might include different details of cosmic ray
acceleration and propagation, or larger magnetic fields at high
redshift. However, it has been shown that magnetic fields of a strength
similar to those found in local galaxies are in place at $z\sim 1$
\citep{bernet08}.

The two redshift bins can also be seen, approximately, as two
luminosity bins. Thus, a different interpretation may be that $q$ is
luminosity dependent, as suggested by \citet{symeonidis2011} that the
X-ray/far-infrared ratio may be lower in UltraLuminous InfraRed
Galaxies (ULIRGs) than in normal galaxies with lower luminosities. If
the analysis presented in this Section is repeated by dividing the
sample in two luminosity bins (radio luminosities lower and higher
than $4\e{29}$ \ergsHz, which is the median luminosity of the SF
sample), then a picture quite similar to Fig.~\ref{fig:qbayes} is
obtained: the high-luminosity bin favours a smaller $q$ than the
low-luminosity bin, yet there is a sizable fraction of the parameter
space which is allowed by both bins, and which contains the RCS03
value.

However, the redshift bins whose data prompt for the evolution only
sample the high-luminosity tail of the radio luminosity function, and
it is possible that a selection bias is part of the explanation: it
may be that radio-luminous objects have lower X-ray luminosity than
average.  In fact, it has been shown \citep{vattakunnel12} that
galaxies in the \chandra\ Deep Fields, whose average X-ray luminosity
is about one order of magnitude lower than those presented here, follow
the same X-ray/radio correlation of galaxies in the local universe.

An estimate of the bias due to the radio selection may be done with
the method employed by \citet{sargent2010} for the infrared/radio
correlation (see also
\citealt{kellerman64,condon84,francis93,lauer07}), in which the bias
for a flux limited survey (where the luminosity function of the
sources is not fully sampled) depends only on the scatter of the
correlation $\sigma$ and on the slope of the differential number
counts $\beta$:
\begin{equation}
  \label{eq:qbias}
  \Delta q_{\mathrm bias}=\mathrm{ln}(10)\, (\beta-1)\, \sigma^2\sim 0.18
\end{equation}
where $\beta\sim 2.35$ (see Sect.~\ref{sec:lognlogs}) and
$\sigma$=0.24. The amount of correction is shown in
Fig.~\ref{fig:qbayes} as the green dashed arrow. Applying this
correction to the $z>0.2$ and any-$z$ bins would mostly remove the
redshift (or luminosity) evolution of $q$. This correction needs not to
be applied to the $z\le 0.2$ bin because here the luminosity function
should be almost correctly sampled. However, this correction should be
only taken as a first-order approximation, because one of its
hypotheses is that the scatter of the X-ray/radio correlation is
described by a Gaussian function. Because the $\Delta q_{\mathrm
  bias}$ is sensitive to the shape of wings of the function, any
deviation of the correlation from a Gaussian would require
Eq.~(\ref{eq:qbias}) to be modified accordingly \citep{lauer07}.

For these reasons, a further investigation of this issue
with deeper observations in both the radio and X-ray bands, and
including a proper statistical treatment of truncated data\footnote{In
  statistical nomenclature, upper limits to the fluxes of objects
  otherwise detected at a different wavelength are an example of
  \textit{censored data}, while a flux-limited survey for which no
  information about the existence of objects at fluxes lower than the
  limit is an example of \textit{truncated data}. Truncated data
  cannot be treated with the same tools valid for censored data
  \citep[see, e.g.,][]{lawless}.} could be an interesting subject for
a follow-up analysis.

\section{Demographics of star forming galaxies}

\begin{figure*}
  \sidecaption
  \centering
  \includegraphics[width=.7\textwidth]{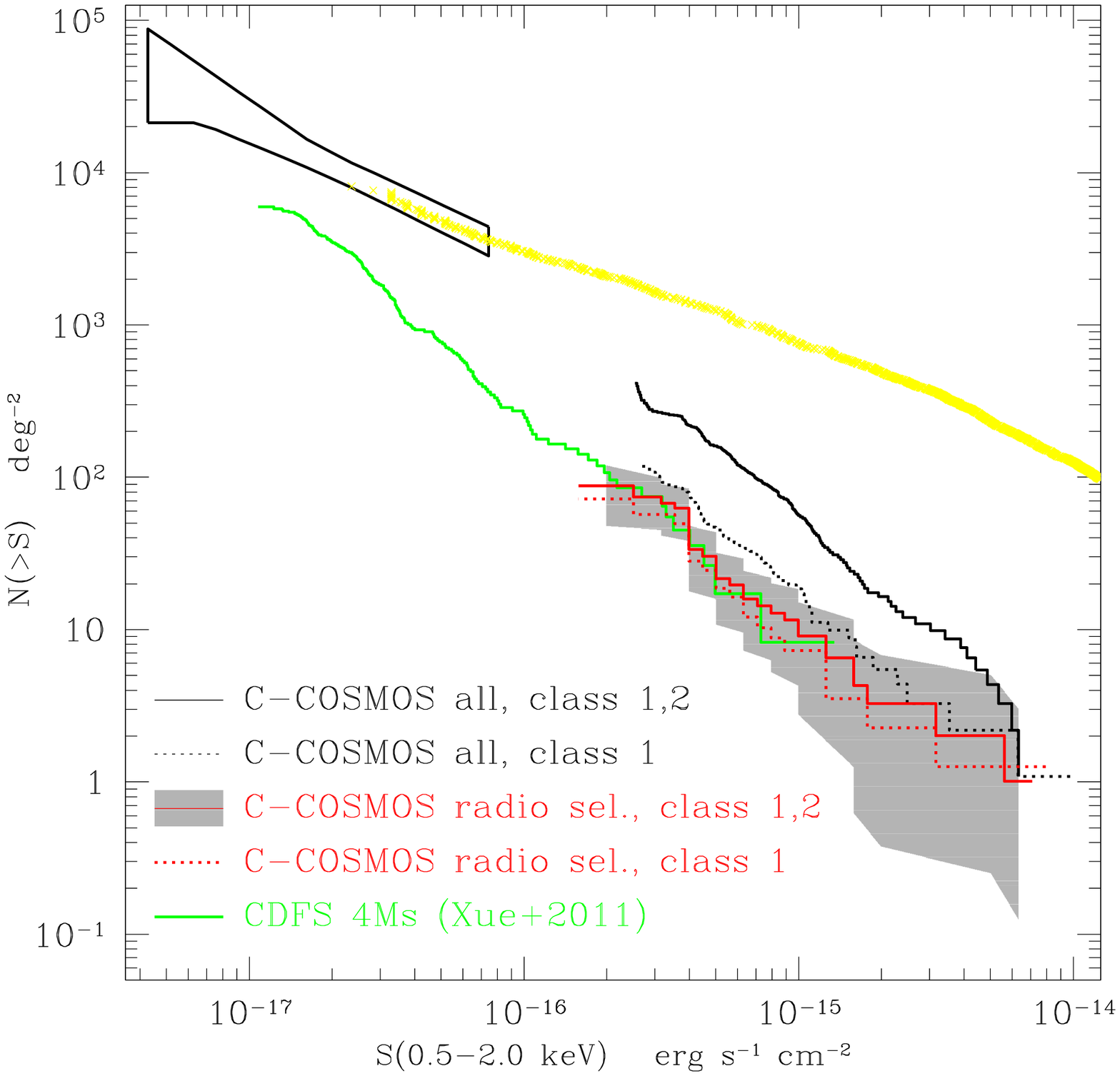}
  \caption{Number counts of SF sources detected in the VLA- and
    C-COSMOS surveys, compared to the CDFS
    determinations. \newline
    Dotted red histogram:
    VLA+C-COSMOS class-1 SF candidates. Solid red histogram, with grey error
    area: VLA+C-COSMOS class-1,2 SF candidates.
    Dotted black histogram: C-COSMOS class-1 SF candidates (not
    requiring radio detection). Solid black histogram:
    C-COSMOS class-1,2 SF candidates (not
    requiring radio detection).
    Green histogram: galaxy candidates in the CDFS
    \citep{xue-cdfs4Ms}; their selection criteria are similar to ours
    for the dotted red histogram.
    The thick yellow line and the horn-shaped
    symbol refer to the total (AGN+SF) number counts and fluctuations
    in the \chandra\ Deep Fields \citep{moretti03,miyaji02a,miyaji02b}.
  }
  \label{fig:lognlogs}
\end{figure*}

\label{sec:lognlogs}

An important test for the selection criteria described so far is to
check for the size of the population of candidate SF X-ray
galaxies. In this Section, we describe four different alternatives,
which are plotted in Fig.~\ref{fig:lognlogs}. The galaxy X-ray number
counts from the \chandra\ Deep Fields \citep[CDFS][]{xue-cdfs4Ms} are also
shown for reference.

First, we consider the 22 class 1 objects from Tables~\ref{tab:cat}
and \ref{tab:catAGN}. This is the strictest selection that we discuss,
since it requires that an object is radio-detected, and that all the
X-ray based criteria are fulfilled. It should thus be considered as a
lower limit. We keep the assumption that a criterion is
considered fulfilled if the relevant data are lacking, as done in
Sect.~\ref{sec:cat}.  The number counts for this selection are plotted
in Fig.~\ref{fig:lognlogs} as the dotted red histogram.

The addition of the 14 class 2 objects to the above selection gives the
solid red histogram, with the errors shown as the grey area. We only
plot the errors for this selection, in order not to clutter the
figure; however, they can be taken as representative of the other
alternatives. The ratio between the class-1-only and the class$\le2$
histograms is about a factor of 2 at high X-ray fluxes and less than
that at lower fluxes.

A different approach is to discard the radio selection criterion, and to apply
the X-ray based criteria to the whole C-COSMOS catalogue\citep{ccosmos-cat,civano12}. This opens
the possibility to include a larger number of composite SF/AGN and of
low luminosity AGN in the selection. A number of 63 class-1 and 192
class-2 objects are selected in this way. The dotted and solid black
histograms show the result for the class-1 and class$\le2$ objects,
respectively. While the class-1 histogram lies a factor of $\sim 2$-3
above the red ones and is still within the 1--2$\sigma$ errors for the
above determinations, a much larger difference (a factor of $\sim 7$)
is observed for the class$\le2$ histogram. 

The latter \lognlogs\ should be regarded as a likely overestimate for
the X-ray galaxy number counts. In fact, if this histogram were
extrapolated at fainter fluxes, it would predict a number of galaxy
which, summed to expected number of AGN from synthesis models
\citep{gilli07,treister09}, would be incompatible with the observed
total \lognlogs.  A further hint comes from the integration of the
galaxy X-ray luminosity function: the theoretical \lognlogs\ relations
in RCS05 would lie on average a factor of 3 below the histogram. (The same
\lognlogs\ are however consistent with the other three
determinations).  

The galaxy number counts were derived by \citet{xue-cdfs4Ms} for the
4Ms CDFS by considering the following criteria: X-ray luminosity,
photon index, X/O, optical spectroscopic classification, and
X-ray/radio ratio. The criteria were joined is a similar manner to
what done here for the class-1 sources. It is thus not surprising
that, although each threshold is somewhat different from what used in
this paper, the final result is very similar to the counts of
radio-selected galaxies presented here. A power-law with the form
$\Log\ N(>S)=-1.35\, \Log\ S -19.15$ may thus be considered a useful
description of both \citet{xue-cdfs4Ms} and our determinations of the
galaxy X-ray number counts.

\section{Conclusions}
\label{sec:conclusions}

We have presented the X-ray properties of a sample of 242 SF galaxies
in the C-COSMOS field, selected in the radio band and classified
according to the optical colours with the method described in
S08. This method builds on the definition of a
synthetic rest-frame colour, which can be calculated from
narrow-band photometry in several bands, and which has been shown to
correlate with the position in the BPT
diagram \citep{smolcic06}. It has a similar power to the BPT diagram,
with the advantage of not requiring expensive spectral
observations.

 In \chandra\ observations, 33 objects were
detected. A comparison sample of 398 candidate type-II AGN (with 82
detections) is also presented.

We have reviewed some X-ray based selection criteria commonly used in
the literature, and analyzed how they affect the composition of the SF
and AGN samples. We have thus refined the SF sample, and recovered
some objects from the AGN one, on the basis of the following parameters:
\begin{itemize}
\item hardness ratio;
\item X-ray luminosity;
\item X-ray/optical flux ratio;
\item classification from optical spectroscopy.
\end{itemize}
This is a small yet effective set of indicators based only on X-ray
and optical properties. We also mention that a similar method has been
applied by \citet{xue-cdfs4Ms} in the \chandra\ Deep Field South.

We have proposed two refined subsamples of C-COSMOS SF galaxies,
based on the absence of AGN-like properties, one (class-1) being more strict in
its criteria than the other (class$\le 2$), containing 8 and
16 objects respectively. If the same method is applied to the AGN sample, 14
objects may be recovered as SF under the stricter method, and 20 under
the more liberal one; these objects may be composite, or may have been
misclassified by the optical colour method.  

Of 33 detections in the SF sample, 17 exhibit AGN-like properties in
terms of hardness ratio, non-detection in the 0.5-2.0 keV band,
optical spectrum, X-ray/optical flux ratio and absolute X-ray
luminosity.  Among 82 detections in the AGN sample, 20 have SF-like
properties. Thus the fraction of composite/misclassified objects is
50\% for the SF sample and 25\% for the AGN, while S08
reported fractions of 30\% and 20\%, respectively.  The larger
fractions observed here are likely to be explained as a selection
effect, due to the fact that AGN are on average brighter than galaxies
in the X-rays.

Conversely, the stacked spectra of X-ray undetected SF and AGN are
significantly different: the SF one can be fit with power-law spectra
with $1.5\lesssim \Gamma\lesssim 2.5$, while the AGN one is flatter
($0.4\lesssim \Gamma\lesssim 0.9$). This suggests that the two samples
do have different physical properties and that the fractions of
composite/misclassified are actually lower for X-ray-undetected
objects.
Thus we regard the fraction of mis-classified and/or composite objects
to be in line with the expectations from S08.

We have investigated if the radio/X-ray luminosities correlation
(RCS03) applies to our data, and if there is any evidence for redshift
evolution of the correlation parameters. A subsample of SF objects at
$z\le 0.2$ yields an X-ray/radio ratio fully consistent with the local
RCS03 estimate. Data at larger redshifts are still consistent with the
local value. Some evolution towards lower X-ray/radio ratios is
possible, but at least part of the evolution may be explained by
selection biases arising in flux-limited surveys. Further analysis of
deeper data, or the use of statistical techniques appropriate to
truncated data may be necessary.

We have presented the number counts of the C-COSMOS SF galaxies 
according to different selection criteria, and compared them to
the number counts of the CDFS galaxies
\citep{xue-cdfs4Ms}. Considering only the radio-selected class-1, or
the radio-selected class$\le 2$, or dropping the radio selection and
only considering class-1, all lead to estimates which are consistent to
each other within the 1--2$\sigma$ errors. Dropping the radio
selection and considering class$\le 2$ objects gives an
overestimate of the galaxy number counts.

Further observations of the COSMOS field with \chandra\ would allow a
much better determination of the X-ray demographics of the SF
galaxies at the redshifts probed in this paper. By extending the
coverage to the full 2 deg$^2$, with a uniform exposure of 180 ks
over 1.7deg$^2$ (the HST-observed area), the final size of the X-ray
detected COSMOS SF galaxy sample should be of the order of 200.

The AEGIS survey \citep{nandra05-aegis} is similar in methodology to
COSMOS, and currently has an observed area and flux limit similar to
C-COSMOS, and recent observations have added deeper coverage
(uniform exposure of 800 ks) to a 0.6 deg$^2$ sub-field.  Even deeper
is the exposure (4 Ms) in the \chandra\ Deep Field South (CDFS) field
in the GOODS survey. Though the probed area is smaller (0.2 deg$^2$;
\citealt{xue-cdfs4Ms,vattakunnel12}), the CDFS already provides 179 objects
classified as galaxies (24\% of the total).
This latter field also has a 3 Ms coverage
with \xmm, which is providing good quality spectroscopy
\citep{comastri11-cdfs}.

The surveys described above have also extensive optical spectroscopy,
and have been observed in the far infrared by {\em Spitzer} and {\em
  Herschel}. The inclusion of infrared data could provide a further
improvement in the object classification, and together with optical
photometry could allow to break down the AGN and host galaxy
contributions for the composite objects.

Finally, these data sets and the classifications done insofar could be
used as testbeds for innovative statistical methods in object
recognition and classification. This would be especially useful in
light of the future large surveys, both in X-rays (e.g., {\em
  eROSITA}) and in optical (LSST, Pan-STARRS, SNAP).

\begin{acknowledgements}
  We thank an anonymous referee whose comments have contributed to
  improve the presentation of this paper.  This research has made use
  of the Perl Data Language (PDL) which provides a high-level
  numerical functionality for the Perl programming language
  \citep[][http://pdl.perl.org]{pdl}.  We acknowledge financial
  contribution from the agreement ASI-INAF I/009/10/0, and 
  a grant from the Greek General Secretariat of Research and
  Technology in the framework of the program Support of Postdoctoral
  Researchers. The research leading to these results has received
  funding from the European Union's Seventh Framework programme under
  grant agreement 229517.
\end{acknowledgements}

\bibliographystyle{aa}
\bibliography{../fullbiblio}

\longtabL{2}{
\begin{landscape}
\begin{longtable}{llrlrrrrrrrlr}
  \caption{Catalogue of radio-selected SF-candidate sources with an
    X-ray detection in C-COSMOS. The columns report: VLA and \chandra\
    names, \chandra\ ID from \citet{ccosmos-cat}, redshifts, X-ray
    fluxes (in \ergscmq) and luminosities (rest-frame; in \ergs) for
    the soft (0.5--2.0 keV) and hard (2.0-10 keV) bands; rest-frame
    hardness ratios; X-ray/optical flux ratios; X-ray/radio flux ratio
    $q$; classification from optical spectroscopy (A: absorption line
    galaxy; E: emission line galaxy; L: spectrum with low signal/noise
    ratio; T1: type-I AGN; T2: type-II AGN); final classification (see
    Sect.~\ref{sec:cat}).  }
  \label{tab:cat}
  \centering
\cr\hline
       VLA name                 &    \chandra\ name     &\chandra &redshift &Flux &Flux     &Lum.          &Lum.      &HR    &X/O      &$q$  &Opt.  &$\!\!\!\!$Class \\  
(COSMOSVLA-)                    &  (CXOC-)              &    ID   &        &\soft &\hard    &\soft         &\hard     &      &ratio    &     &class.&      \\            
\hline
J095844.86+021100.3 & J095844.8+021100 & 327   &0.633   &$ 7.27\e{-16}$ &$ 2.98\e{-15}$ &$ 1.24\e{42}$ &$ 5.08\e{42}$ &-0.03 &-0.78    &11.74    &E  &3   \\     
J095903.81+020316.8 & J095903.8+020317 & 1065  &1.247   &$ 2.20\e{-16}$ &$<2.32\e{-15}$ &$ 1.99\e{42}$ &$<2.09\e{43}$ & 0.14 &-0.83    &11.19    &T1 &3   \\     %
J095915.63+020856.6 & J095915.6+020857 & 2800  &0.168   &$ 4.57\e{-16}$ &$<1.82\e{-15}$ &$ 3.60\e{40}$ &$<1.43\e{41}$ &-0.52 &-2.07    &11.41    &T2 &2   \\     %
J095924.03+022708.1 & J095924.0+022708 & 2247  &0.714   &$<8.24\e{-16}$ &$ 4.39\e{-15}$ &$<1.89\e{42}$ &$ 1.01\e{43}$ & 0.34 &$<$-0.33 &$<$10.86 &E &3   \\     
J095934.70+021228.9 & J095934.7+021229 & 1542  &0.345   &$ 4.18\e{-16}$ &$<2.18\e{-15}$ &$ 1.68\e{41}$ &$<8.76\e{41}$ &-0.22 &-1.59    &10.81    &E &1   \\     
J095941.06+021501.3 & J095941.0+021501 & 1295  &0.880   &$<3.13\e{-16}$ &$ 4.53\e{-15}$ &$<1.39\e{42}$ &$ 2.01\e{43}$ & 0.68 &$<$-0.78 &$<$11.67 &E &3   \\     %
J095945.19+023439.4 & J095945.2+023439 & 1272  &0.124   &$ 4.34\e{-16}$ &$ 4.46\e{-15}$ &$ 1.75\e{40}$ &$ 1.79\e{41}$ & 0.64 &-2.29    &11.13    &L  &2   \\     %
J095947.61+015555.9 & J095947.6+015555 & 811   &0.27$^a$&$ 4.28\e{-16}$ &$<1.71\e{-15}$ &$ 1.56\e{41}$ &$<6.23\e{41}$ &-0.46 &-1.55    &11.37    &--- &1   \\     %
J095953.91+024043.3 & J095953.9+024043 & 612   &0.537   &$<6.42\e{-16}$ &$ 1.35\e{-14}$ &$<3.05\e{41}$ &$ 6.42\e{42}$ & 0.88 &$<$-0.53 &$<$11.73 &E &3   \\     %
J100002.06+020132.6 & J100002.0+020132 & 1499  &1.193   &$<4.33\e{-16}$ &$ 3.98\e{-15}$ &$<3.50\e{42}$ &$ 3.22\e{43}$ & 0.26 &$<$-0.15 &$<$11.73 &E &3   \\     
J100004.53+020852.2 & J100004.5+020852 & 1126  &0.959   &$ 8.65\e{-16}$ &$ 1.66\e{-14}$ &$ 4.08\e{42}$ &$ 7.83\e{43}$ & 0.54 &-0.43    &11.84    &E &3   \\     
J100005.31+020135.3 & J100005.4+020134 & 1083  &0.311   &$ 3.96\e{-16}$ &$ 2.84\e{-15}$ &$ 1.25\e{41}$ &$ 8.97\e{41}$ & 0.62 &-1.66    &11.43    &E &2   \\     
J100010.15+024141.4 & J100010.2+024140 & 623   &0.221   &$ 1.50\e{-15}$ &$ 7.59\e{-15}$ &$ 1.21\e{41}$ &$ 6.12\e{41}$ & 0.07 &-1.43    &10.75    &T2 &2   \\     %
J100011.90+021425.1 & J100011.9+021425 & 1517  &0.602   &$<1.48\e{-16}$ &$ 3.68\e{-15}$ &$<2.23\e{41}$ &$ 5.55\e{42}$ & 0.95 &$<$-1.10 &$<$11.36 &T2 &3   \\     
J100013.58+021230.8 & J100013.5+021230 & 1164  &0.188   &$ 3.18\e{-16}$ &$<2.89\e{-15}$ &$ 1.14\e{40}$ &$<1.03\e{41}$ & 0.26 &-1.92    &10.98    &E &2  \\     %
J100013.72+021221.4 & J100013.7+021221 & 1297  &0.186   &$ 5.97\e{-16}$ &$ 5.40\e{-15}$ &$ 5.91\e{40}$ &$ 5.34\e{41}$ & 0.25 &-1.82    &11.21    &T2 &3   \\     %
J100022.94+021312.5 & J100022.9+021313 & 1221  &0.186   &$ 6.85\e{-16}$ &$<1.15\e{-15}$ &$ 6.74\e{40}$ &$<1.13\e{41}$ &-0.76 &-2.00    &11.37    &E &1   \\     
J100023.82+020105.4 & J100023.8+020106 & 1092  &0.361   &$ 4.08\e{-16}$ &$<2.59\e{-15}$ &$ 1.89\e{41}$ &$<1.20\e{42}$ & 0.08 &-1.64    &11.08    &E &1   \\     %
J100027.77+015704.2 & J100027.7+015705 & 1096  &0.26$^a$&$ 3.83\e{-16}$ &$<1.03\e{-15}$ &$ 8.33\e{40}$ &$<2.24\e{41}$ &-0.77 &-1.50    &11.62    &--- &2   \\     
J100028.55+022725.9 & J100028.5+022725 & 1535  &0.248   &$ 4.42\e{-16}$ &$<1.14\e{-15}$ &$ 8.22\e{40}$ &$<2.12\e{41}$ &-0.20 &-1.69    &10.82    &E &1   \\     %
J100045.10+022110.5 & J100045.1+022110 & 962   &0.265   &$ 1.03\e{-15}$ &$ 4.04\e{-15}$ &$ 2.25\e{41}$ &$ 8.84\e{41}$ & 0.03 &-0.85    &11.97    &T2 &3   \\     %
J100051.34+021117.7 & J100051.3+021117 & 12650 &0.802   &$ 2.06\e{-16}$ &$<1.32\e{-15}$ &$ 6.24\e{41}$ &$<4.00\e{42}$ &-0.17 &-1.16    &11.42    &E &1   \\     
J100054.74+021611.3 & J100054.7+021611 & 24    &0.678   &$ 1.33\e{-15}$ &$ 5.23\e{-15}$ &$ 2.68\e{42}$ &$ 1.05\e{43}$ &-0.12 &-0.41    &11.85    &E &3   \\     
J100056.16+015642.7 & J100056.1+015642 & 366   &0.360   &$ 3.16\e{-15}$ &$ 6.57\e{-15}$ &$ 1.40\e{42}$ &$ 2.91\e{42}$ &-0.44 &-0.72    &12.39    &T2 &3   \\     %
J100057.46+015901.6 & J100057.4+015902 & 368   &0.511   &$<1.70\e{-16}$ &$ 6.22\e{-15}$ &$<1.73\e{41}$ &$ 6.34\e{42}$ & 0.98 &$<$-1.42 &$<$11.39 &T2 &3   \\     
J100058.69+021139.6 & J100058.6+021139 & 11100 &0.110   &$ 3.66\e{-16}$ &$<1.00\e{-15}$ &$ 1.13\e{40}$ &$<3.09\e{40}$ &-0.89 &-2.21    &11.39    &E &1   \\     
J100059.70+015820.2 & J100059.6+015820 & 1508  &0.674   &$<2.65\e{-16}$ &$ 1.65\e{-15}$ &$<5.25\e{41}$ &$ 3.27\e{42}$ & 0.07 &$<$-1.20 &$<$11.65 &E &2   \\     
J100100.67+021641.4 & J100100.7+021640 & 3718  &0.166   &$ 9.46\e{-16}$ &$<2.00\e{-15}$ &$ 7.20\e{40}$ &$<1.52\e{41}$ &-0.32 &-1.76    &11.27    &T2 &2   \\     %
J100101.94+014800.3 & J100101.9+014800 & 284   &0.909   &$<1.40\e{-16}$ &$ 9.19\e{-15}$ &$<5.76\e{41}$ &$ 3.78\e{43}$ & 0.96 &$<$-1.24 &$<$11.03 &E &3   \\     %
J100114.96+014348.4 & J100114.9+014348 & 298   &0.578   &$<4.30\e{-16}$ &$ 1.25\e{-14}$ &$<5.93\e{41}$ &$ 1.72\e{43}$ & 0.80 &$<$-1.13 &$<$11.42 &E &3   \\     
J100129.41+013633.4 & J100129.3+013634 & 1678  &0.104   &$ 3.61\e{-14}$ &$ 6.34\e{-13}$ &$ 1.00\e{42}$ &$ 1.76\e{43}$ & 0.55 &-0.62    &13.01    &E &3   \\     %
J100129.95+021705.2 & J100129.9+021705 & 2078  &0.076   &$ 2.96\e{-16}$ &$<1.13\e{-15}$ &$ 4.16\e{39}$ &$<1.59\e{40}$ &-0.55 &-2.45    &10.98    &E &1   \\     %
J100203.09+020907.1 & J100203.1+020906 & 551   &0.556   &$<2.74\e{-16}$ &$ 6.29\e{-15}$ &$<3.40\e{41}$ &$ 7.81\e{42}$ & 0.94 &$<$-1.42 &$<$11.45 &E &3   \\     
\hline
\renewcommand*\footnoterule{}
\footnotetext[1]{Photometric redshift.}

\end{longtable}
\end{landscape}
}

\longtabL{3}{
\begin{landscape}
\begin{longtable}{llrlrrrrrrrlr}
  \caption{Catalogue of radio-selected AGN-candidate sources with an
    X-ray detection in C-COSMOS. Columns and footnotes as in Table\ref{tab:cat}. 
}
  \label{tab:catAGN}
  \centering
\cr\hline
       VLA name       &    \chandra\ name& \chandra &redshift &Flux  &Flux  &Lum.  &Lum.  &HR  &X/O   &$q$  &Opt.  &$\!\!\!\!$Class \\  
  (COSMOSVLA-)        & (CXOC-)          &    ID    &         &\soft &\hard &\soft &\hard &    &ratio &     &class.&      \\            
\hline
  J095833.84+020627.8 & J095833.8+020628 & 3570  &1.244   & $<2.50\e{-16}$ &$ 5.22\e{-15}$ &$ <2.43\e{42}$ &$  5.07\e{43}$& 0.95 & -0.81& 11.57&E  &3\\
  J095835.71+021357.9 & J095835.6+021358 & 2848  &1.27$^a$& $ 1.01\e{-15}$ &$ 5.14\e{-15}$ &$  3.10\e{42}$ &$  1.58\e{43}$&-0.16 & 0.34 &11.56 &---&3\\
  J095835.91+021233.3 & J095835.8+021234 & 680   &0.954   & $ 7.58\e{-16}$ &$<2.98\e{-15}$ &$  3.77\e{42}$ &$ <1.48\e{43}$& 0.24 & 0.02 &11.26 &E  &3\\
  J095844.72+020249.7 & J095844.7+020250 & 1001  &0.093   & $ 1.12\e{-15}$ &$<1.18\e{-15}$ &$  2.47\e{40}$ &$ <2.60\e{40}$&-0.90 & 0.19 &11.05 &E  &1\\ 
  J095846.01+014905.5 & J095846.0+014905 & 196   &0.738   & $ 8.39\e{-16}$ &$ 6.30\e{-15}$ &$  2.20\e{42}$ &$  1.65\e{43}$& 0.14 & 0.24 &12.05 &E  &3\\
  J095847.00+021552.2 & J095847.0+021552 & 428   &0.552   & $ 1.27\e{-14}$ &$ 7.17\e{-14}$ &$  1.62\e{43}$ &$  9.17\e{43}$& 0.04 & 1.60 &12.94 &E  &3\\
  J095857.03+020354.9 & J095857.0+020355 & 163   &0.678   & $ 4.63\e{-15}$ &$ 1.22\e{-14}$ &$  9.83\e{42}$ &$  2.59\e{43}$&-0.34 & 0.93 &12.86 &T2 &3\\
  J095857.22+015843.6 & J095857.2+015843 & 164   &0.527   & $ 7.12\e{-16}$ &$ 3.02\e{-14}$ &$  8.13\e{41}$ &$  3.45\e{43}$& 0.76 & -0.46& 12.02&E  &3\\
  J095858.53+021459.1 & J095858.5+021459 & 418   &0.132   & $ 5.53\e{-14}$ &$ 1.47\e{-13}$ &$  2.59\e{42}$ &$  6.90\e{42}$&-0.30 & 1.17 &12.58 &T1 &3\\
  J095910.31+020732.4 & J095910.3+020732 & 679   &0.353   & $ 1.43\e{-15}$ &$<2.54\e{-15}$ &$  6.24\e{41}$ &$ <1.11\e{42}$&-0.70 & 0.66 &11.58 &T2 &2\\
  J095916.52+020944.4 & J095916.5+020944 & 827   &0.354   & $ 8.86\e{-16}$ &$<1.59\e{-15}$ &$  3.89\e{41}$ &$ <6.99\e{41}$&-0.92 & 0.29 &11.43 &L  &1\\ 
  J095920.17+021831.3 & J095920.1+021831 & 411   &1.162   & $ 6.41\e{-15}$ &$ 2.11\e{-14}$ &$  5.25\e{43}$ &$  1.73\e{44}$&-0.23 & -0.30& 13.11&T2 &3\\
  J095920.98+015818.6 & J095921.0+015818 & 1467  &0.89$^a$& $ 3.30\e{-16}$ &$ 2.22\e{-15}$ &$  1.84\e{42}$ &$  1.24\e{43}$&-0.04 & -1.21& 11.76&---&3\\
  J095921.37+022427.9 & J095921.3+022427 & 1189  &0.732   & $ 1.02\e{-15}$ &$ 1.89\e{-15}$ &$  2.62\e{42}$ &$  4.85\e{42}$&-0.84 & -0.98& 12.11&E  &3\\
  J095921.85+021517.5 & J095921.8+021518 & 856   &0.345   & $<2.69\e{-16}$ &$ 2.97\e{-15}$ &$ <1.11\e{41}$ &$  1.23\e{42}$& 0.66 & -2.19& 11.61&E  &2\\
  J095922.18+021927.7 & J095922.2+021927 & 1119  &0.72$^a$& $<2.76\e{-16}$ &$ 5.60\e{-15}$ &$ <2.50\e{42}$ &$  5.08\e{43}$& 0.77 & -1.25& 11.75&---&3\\
  J095925.62+021623.1 & J095925.6+021624 & 2454  &0.764   & $<1.49\e{-16}$ &$ 2.38\e{-15}$ &$ <4.25\e{41}$ &$  6.80\e{42}$& 0.90 & -1.74& 11.31&T2 &3\\
  J095926.83+021023.2 & J095926.8+021023 & 828   &0.884   & $ 3.11\e{-16}$ &$ 2.72\e{-15}$ &$  1.28\e{42}$ &$  1.12\e{43}$& 0.53 & -1.04& 11.42&T2 &3\\
  J095926.89+015341.2 & J095926.8+015341 & 209   &0.445   & $ 1.20\e{-14}$ &$ 5.46\e{-14}$ &$  9.10\e{42}$ &$  4.14\e{43}$&-0.07 & -0.86& 13.05&T2 &3\\
  J095927.27+014634.8 & J095927.3+014636 & 1428  &0.427   & $ 1.81\e{-15}$ &$ 8.17\e{-15}$ &$  1.24\e{42}$ &$  5.61\e{42}$&-0.16 & -0.46& 12.47&T2 &3\\
  J095928.14+020622.9 & J095928.1+020623 & 335   &1.051   & $ 1.85\e{-15}$ &$ 1.17\e{-14}$ &$  1.18\e{43}$ &$  7.43\e{43}$& 0.09 & -0.96& 12.48&E  &3\\
  J095928.18+014451.6 & J095928.0+014452 & 3432  &0.46$^a$& $<4.33\e{-16}$ &$ 7.85\e{-15}$ &$ <5.81\e{41}$ &$  1.05\e{43}$& 0.88 & -0.93& 11.23&---&3\\
  J095931.53+015306.6 & J095931.5+015306 & 3439  &0.531   & $<2.27\e{-16}$ &$ 4.15\e{-15}$ &$ <2.64\e{41}$ &$  4.83\e{42}$& 0.89 & -1.66& 11.63&E  &3\\
  J095932.49+021037.6 & J095932.4+021037 & 682   &0.66    & $<2.62\e{-16}$ &$ 1.84\e{-14}$ &$ <5.20\e{41}$ &$  3.65\e{43}$& 0.84 & -0.16& 10.75&E  &3\\
  J095934.46+020628.3 & J095934.4+020628 & 339   &0.686   & $ 1.01\e{-14}$ &$ 2.77\e{-14}$ &$  2.21\e{43}$ &$  6.05\e{43}$&-0.30 & 0.73 &13.04 &E  &3\\
  J095934.64+015650.9 & J095934.6+015651 & 1711  &0.772   & $ 8.53\e{-16}$ &$<2.26\e{-15}$ &$  2.50\e{42}$ &$ <6.62\e{42}$&-0.50 & -0.93& 12.00&E  &3\\
  J095937.41+022347.3 & J095937.4+022347 & 701   &0.74    & $ 9.35\e{-16}$ &$ 2.21\e{-15}$ &$  2.47\e{42}$ &$  5.83\e{42}$&-0.34 & -0.94& 10.87&E  &3\\
  J095939.77+023116.9 & J095939.8+023117 & 1310  &0.729   & $ 9.76\e{-16}$ &$ 2.71\e{-15}$ &$  2.48\e{42}$ &$  6.89\e{42}$&-0.43 & -0.19& 11.68&A  &3\\ 
  J095942.25+023337.8 & J095942.3+023338 & 1414  &0.701   & $<1.53\e{-16}$ &$ 2.89\e{-15}$ &$ <3.53\e{41}$ &$  6.66\e{42}$& 0.94 & -1.78& 11.41&E  &3\\ 
  J095942.72+023206.5 & J095942.6+023205 & 1231  &0.668   & $<3.39\e{-16}$ &$ 4.36\e{-15}$ &$ <6.93\e{41}$ &$  8.92\e{42}$& 0.90 & -1.99& 11.21&A  &3\\ 
  J095948.82+021245.1 & J095948.7+021245 & 1128  &0.478   & $ 5.45\e{-16}$ &$ 3.06\e{-15}$ &$  4.91\e{41}$ &$  2.76\e{42}$&-0.12 & -0.73& 11.27&A  &1\\ 
  J095950.26+014805.3 & J095950.2+014805 & 669   &0.68    & $<3.23\e{-16}$ &$ 3.79\e{-15}$ &$ <6.90\e{41}$ &$  8.10\e{42}$& 0.90 & -0.49& 10.43&E  &3\\
  J095954.68+023015.3 & J095954.6+023015 & 1242  &1.26$^a$& $ 4.81\e{-16}$ &$ 3.71\e{-15}$ &$  4.21\e{42}$ &$  3.24\e{43}$& 0.11 & -0.78& 10.56&---&3\\
  J095956.03+014727.7 & J095956.0+014727 & 246   &0.337   & $ 6.45\e{-15}$ &$ 1.04\e{-14}$ &$  2.52\e{42}$ &$  4.07\e{42}$&-0.48 & -0.46& 12.67&E  &3\\
  J095958.26+015128.1 & J095958.2+015128 & 233   &0.834   & $ 2.33\e{-15}$ &$ 2.74\e{-15}$ &$  8.28\e{42}$ &$  9.74\e{42}$&-0.57 & -1.08& 12.65&---&3\\
  J100005.36+023059.5 & J100005.3+023059 & 105   &0.68    & $ 3.33\e{-15}$ &$ 2.00\e{-14}$ &$  7.12\e{42}$ &$  4.28\e{43}$& 0.14 & -1.81& 12.01&E  &3\\
  J100005.99+015453.4 & J100006.0+015453 & 668   &0.969   & $ 5.35\e{-16}$ &$ 2.12\e{-15}$ &$  2.77\e{42}$ &$  1.10\e{43}$&-0.06 & -2.14& 11.62&E  &3\\
  J100008.43+020247.3 & J100008.5+020247 & 1089  &0.37    & $ 3.73\e{-16}$ &$ 2.31\e{-15}$ &$  1.82\e{41}$ &$  1.13\e{42}$& 0.17 & -1.92& 11.34&T2 &3\\
  J100012.06+014440.1 & J100012.0+014440 & 266   &1.148   & $ 4.08\e{-15}$ &$ 5.95\e{-15}$ &$  3.24\e{43}$ &$  4.72\e{43}$&-0.59 & -0.58& 12.85&T1 &3\\
  J100013.93+022249.6 & J100013.9+022249 & 898   &0.347   & $ 1.62\e{-15}$ &$<1.11\e{-15}$ &$  6.79\e{41}$ &$ <4.65\e{41}$&-0.89 & -1.15& 12.11&A  &1\\ 
  J100014.19+021312.1 & J100014.1+021311 & 1139  &1.141   & $ 2.45\e{-15}$ &$ 4.18\e{-15}$ &$  1.91\e{43}$ &$  3.27\e{43}$&-0.51 & -1.79& 11.35&T2 &3\\
\caption{continued.}                                                                                                                                                
\cr\hline                                                                                                                                                             
       VLA name                  &    \chandra\ name    & \chandra &redshift &Flux  &Flux  &Lum.  &Lum.  &HR  &X/O   &$q$  &Opt.  &$\!\!\!\!$Class \\  
 (COSMOSVLA-)                    &   (CXOC-)            &    ID    &         &\soft &\hard &\soft &\hard &    &ratio &     &class.&      \\            
\hline                                                                                                                                                             
  J100014.27+020644.4 & J100014.2+020644 & 1082  &1.02$^a$& $ 9.54\e{-16}$ &$ 6.94\e{-15}$ &$  8.20\e{42}$ &$  5.97\e{43}$ &  0.29 & -0.88& 12.06&---&3\\
  J100016.05+021237.4 & J100016.0+021237 & 460   &0.187   & $<4.74\e{-16}$ &$ 8.22\e{-15}$ &$ <4.80\e{40}$ &$  8.33\e{41}$ &  0.76 & -2.22& 11.02&E  &2\\
  J100019.32+022324.9 & J100019.1+022324 & 3673  &0.353   & $<3.85\e{-16}$ &$ 4.13\e{-15}$ &$ <1.68\e{41}$ &$  1.80\e{42}$ &  0.53 & -2.59& 11.62&T2 &3\\
  J100021.42+021758.4 & J100021.4+021759 & 2125  &0.533   & $<8.74\e{-17}$ &$ 3.43\e{-15}$ &$ <1.03\e{41}$ &$  4.03\e{42}$ &  0.98 & -2.71& 11.06&T2 &3\\
  J100021.78+020000.2 & J100021.7+020000 & 1289  &0.219   & $ 1.06\e{-15}$ &$ 2.98\e{-15}$ &$  1.53\e{41}$ &$  4.31\e{41}$ & -0.18 & -2.00& 10.89&L  &1\\ 
  J100030.73+014711.0 & J100030.7+014711 & 1036  &1.42$^a$& $ 1.53\e{-15}$ &$ 1.34\e{-14}$ &$  1.41\e{43}$ &$  1.23\e{44}$ & 0.22  & 0.04 &12.09 &---&3\\
  J100036.05+022830.6 & J100036.0+022830 & 77    &0.688   & $ 2.83\e{-15}$ &$ 4.31\e{-14}$ &$  6.23\e{42}$ &$  9.48\e{43}$ & 0.50  & 0.52 &11.81 &E  &3\\
  J100037.65+022949.0 & J100037.6+022949 & 3648  &0.671   & $<3.38\e{-16}$ &$<2.77\e{-15}$ &$ <6.99\e{41}$ &$ <5.73\e{42}$ & 0.45  & 0.01 &11.32 &T2 &3\\
  J100043.14+020637.1 & J100043.1+020637 & 42    &0.36    & $ 1.82\e{-14}$ &$ 3.62\e{-14}$ &$  8.32\e{42}$ &$  1.65\e{43}$ &-0.39  & 1.15 &13.17 &T1 &3\\
  J100043.17+014607.9 & J100043.1+014608 & 288   &0.346   & $ 2.13\e{-15}$ &$<2.29\e{-15}$ &$  8.87\e{41}$ &$ <9.53\e{41}$ &-0.94  & -0.35& 9.47 &A  &1\\ 
  J100043.53+022524.5 & J100043.5+022524 & 80    &0.729   & $ 2.61\e{-16}$ &$ 4.34\e{-15}$ &$  6.63\e{41}$ &$  1.10\e{43}$ &  0.44 & -1.49& 10.74&E  &3\\
  J100047.60+015910.3 & J100047.5+015910 & 364   &0.438   & $ 6.50\e{-16}$ &$<2.43\e{-15}$ &$  4.74\e{41}$ &$ <1.77\e{42}$ &-0.47  & -0.77& 9.58 &E  &1\\ 
  J100049.58+014923.7 & J100049.6+014924 & 1292  &0.530   & $ 4.52\e{-16}$ &$<2.20\e{-15}$ &$  5.23\e{41}$ &$ <2.55\e{42}$ &-0.23  & -0.74& 9.71 &L  &1\\ 
  J100049.65+014048.9 & J100049.6+014049 & 294   &1.08$^a$& $ 8.07\e{-16}$ &$ 1.13\e{-14}$ &$  1.94\e{42}$ &$  2.72\e{43}$ &  0.51 & -0.29& 11.58&---&3\\
  J100054.79+014602.5 & J100054.7+014602 & 1049  &0.35    & $ 4.63\e{-16}$ &$ 9.53\e{-15}$ &$  1.98\e{41}$ &$  4.08\e{42}$ &  0.44 & -0.69& 11.61&T2 &3\\
  J100058.20+014559.0 & J100058.1+014559 & 281   &0.623   & $ 2.41\e{-14}$ &$ 8.46\e{-14}$ &$  4.15\e{43}$ &$  1.46\e{44}$ &-0.19  & 1.07 &13.37 &T2 &3\\
  J100100.04+021252.5 & J100100.0+021252 & 28    &0.497   & $ 6.76\e{-16}$ &$ 1.38\e{-14}$ &$  6.70\e{41}$ &$  1.37\e{43}$ &  0.58 & -0.74& 11.74&E  &3\\
  J100103.94+020259.6 & J100103.9+020259 & 928   &0.962   & $ 2.76\e{-16}$ &$ 3.01\e{-15}$ &$  1.40\e{42}$ &$  1.53\e{43}$ &  0.48 & -1.33& 11.40&A  &3\\ 
  J100104.26+023307.4 & J100104.2+023307 & 1871  &0.267   & $ 5.79\e{-16}$ &$<3.19\e{-15}$ &$  1.32\e{41}$ &$ <7.26\e{41}$ & -0.41 & -0.18& 11.90&E  &1\\
  J100107.31+021100.4 & J100107.3+021100 & 1450  &1.241   & $ 3.56\e{-16}$ &$<1.54\e{-15}$ &$  3.44\e{42}$ &$ <1.49\e{43}$ & -0.36 & -0.78& 11.54&T2 &3\\
  J100110.75+020204.2 & J100110.7+020205 & 763   &0.972   & $<2.89\e{-16}$ &$ 3.62\e{-15}$ &$ <1.51\e{42}$ &$  1.89\e{43}$ &  0.90 & -1.01& 10.85&T2 &3\\
  J100111.68+021250.7 & J100111.6+021250 & 760   &1.147   & $ 5.89\e{-16}$ &$ 7.09\e{-15}$ &$  4.66\e{42}$ &$  5.61\e{43}$ &  0.32 & -0.83& 11.79&T2 &3\\
  J100114.85+020208.8 & J100114.8+020208 & 54    &0.971   & $ 1.32\e{-14}$ &$ 3.28\e{-14}$ &$  6.87\e{43}$ &$  1.71\e{44}$ & -0.34 & -0.51& 11.53&T1 &3\\
  J100119.57+015516.3 & J100119.5+015516 & 1368  &0.099   & $ 8.11\e{-16}$ &$<1.15\e{-15}$ &$  2.04\e{40}$ &$ <2.90\e{40}$ & -0.88 & -1.59& 10.80&E  &1\\ 
  J100131.14+022924.7 & J100131.1+022924 & 633   &0.35    & $ 6.32\e{-15}$ &$ 5.21\e{-15}$ &$  2.70\e{42}$ &$  2.23\e{42}$ &-0.78  & 0.51 &11.26 &E   &2\\ 
  J100136.46+022641.8 & J100136.4+022642 & 1811  &0.123   & $ 5.38\e{-16}$ &$<1.86\e{-15}$ &$  2.16\e{40}$ &$ <7.48\e{40}$ & -0.40 & -1.12& 10.60&A  &1\\ 
  J100139.76+022548.8 & J100139.7+022548 & 634   &0.124   & $ 5.01\e{-15}$ &$ 3.50\e{-15}$ &$  2.05\e{41}$ &$  1.43\e{41}$ &-0.63  & 0.29 &11.97 &L  &1\\ 
  J100140.38+020506.7 & J100140.3+020506 & 478   &0.425   & $ 9.09\e{-15}$ &$ 2.62\e{-14}$ &$  6.17\e{42}$ &$  1.78\e{43}$ &-0.25  & 0.55 &13.12 &T2 &3\\
  J100141.03+015904.0 & J100141.0+015906 & 2598  &0.31$^a$& $<3.34\e{-16}$ &$ 3.72\e{-15}$ &$ <1.24\e{41}$ &$  1.39\e{42}$ &  0.68 & -0.53& 10.74&---&2\\
  J100141.32+021031.3 & J100141.3+021031 & 69    &0.983   & $ 2.03\e{-14}$ &$ 3.74\e{-14}$ &$  1.09\e{44}$ &$  2.01\e{44}$ &-0.42  & 1.18 &13.41 &T1 &3\\
  J100141.97+020358.3 & J100141.9+020358 & 482   &0.125   & $<5.86\e{-16}$ &$ 2.09\e{-14}$ &$ <2.44\e{40}$ &$  8.71\e{41}$ &  0.88 & -0.42& 11.67&L  &2\\ 
  J100147.28+015258.2 & J100147.2+015258 & 690   &0.66$^a$& $ 1.18\e{-15}$ &$<2.54\e{-15}$ &$  2.70\e{42}$ &$ <5.81\e{42}$ & -0.81 & -0.66& 11.19&---&3\\
  J100147.35+020314.2 & J100147.4+020314 & 2605  &0.323   & $ 1.12\e{-15}$ &$<2.97\e{-15}$ &$  3.97\e{41}$ &$ <1.05\e{42}$ & -0.62 & -0.17& 10.23&L  &1\\ 
  J100151.21+020715.0 & J100151.2+020714 & 11725 &0.41$^a$& $ 4.32\e{-16}$ &$<9.46\e{-16}$ &$  2.26\e{41}$ &$ <4.95\e{41}$ & -0.57 & -0.06& 11.77&---&1\\
  J100152.22+015608.6 & J100152.1+015608 & 401   &0.969   & $<6.54\e{-16}$ &$ 1.32\e{-14}$ &$ <3.39\e{42}$ &$  6.84\e{43}$ &  0.48 & -0.18& 11.31&T2 &3\\
  J100157.92+021518.9 & J100157.9+021518 & 2016  &0.742   & $ 4.47\e{-16}$ &$<2.75\e{-15}$ &$  1.19\e{42}$ &$ <7.30\e{42}$ & -0.12 & -1.09& 11.87&T2 &3\\
  J100159.16+020521.3 & J100159.1+020522 & 2327  &1.189   & $ 1.43\e{-15}$ &$ 6.66\e{-15}$ &$  1.24\e{43}$ &$  5.78\e{43}$ & -0.02 & -0.30& 12.16&E  &3\\
  J100202.54+020145.3 & J100202.5+020145 & 496   &0.899   & $ 6.53\e{-15}$ &$ 2.49\e{-14}$ &$  2.80\e{43}$ &$  1.07\e{44}$ & -0.11 & -0.46& 12.34&E  &3\\
  J100223.02+022009.8 & J100223.0+022009 & 1769  &0.95$^a$& $ 3.05\e{-15}$ &$ 1.46\e{-14}$ &$  2.24\e{43}$ &$  1.07\e{44}$ & -0.16 & -0.17& 12.36&---&3\\
  J100225.33+022614.1 & J100225.3+022614 & 1574  &1.32$^a$& $<1.01\e{-15}$ &$ 1.17\e{-14}$ &$ <5.53\e{42}$ &$  6.41\e{43}$ &  0.64 & -0.05& 11.85&---&3\\
  J100233.35+022751.9 & J100233.3+022751 & 1586  &0.8     & $ 2.96\e{-14}$ &$ 3.85\e{-14}$ &$  9.48\e{43}$ &$  1.23\e{44}$ &-0.21  & 1.03 &13.27 &E  &3\\
\hline

\end{longtable}
\end{landscape}
}

\end{document}